\documentclass[reprint,nofootinbib,prd]{revtex4-2}

\usepackage{graphicx}
\usepackage{graphics}
\usepackage{dcolumn}
\usepackage{multirow}
\usepackage[hidelinks]{hyperref}

\usepackage{amsfonts}
\usepackage{amsmath}
\usepackage{mathtools}
\DeclareMathAlphabet{\mathcal}{OMS}{cmsy}{m}{n}

\newcommand{\Tr}[1]{\text{Tr}\left\{#1\right\}}

\newcommand{\Nc}{{N_c}}
\newcommand{\Nf}{{N_f}}
\newcommand{\psibar}{{\overline{\psi}}}

\newcommand{\cO}{{\cal O}}
\newcommand{\cZ}{{\cal Z}}
\newcommand{\MSbar}{{\overline{\text{MS}}}}
\newcommand{\corr}[1]{{\langle#1\rangle}}
\newcommand{\corrno}[1]{{\langle#1\rangle}} 
\newcommand{\bs}[1]{{\boldsymbol{#1}}}
\renewcommand{\vec}[1]{{\boldsymbol{#1}}}

\newcommand{\vxi}{{\bs \xi}}

\newcommand{\mt}{{\widetilde m}}

\newcommand{\etabar}{{\overline \eta}}

\newcommand{\cF}{{\cal F}}
\newcommand{\rmc}{{\rm c}}

\newcommand{\albe}{{\alpha\beta}}

\newcommand{\mtht}[1]{{\texorpdfstring{#1}{}}}

\newcommand{\wt}[1]{{\widetilde #1}}

\newcommand{\cn}{{{\rm c}_\nu}}
\newcommand{\mcr}{{m_{\rm cr}}}
\newcommand{\rmcr}{{\rm cr}}

\newcommand{\phat}{{\hat p}}
\newcommand{\pbar}{{\overline p}}

\newcommand{\bsv}{{\bs{v}}}

\newcommand{\static}{{\frac{\Delta f_\vxi^\infty}{\Delta\xi_k}}}
\newcommand{\dynamic}{{\frac{\Delta (f_\vxi-f_\vxi^\infty)}{\Delta\xi_k}}}

\def\nab#1{{\nabla_{#1}}}
\def\nabstar#1{{\nabla\kern0.5pt\smash{\raise 4.5pt\hbox{$\ast$}}
               \kern-5.5pt_{#1}}}

\usepackage{xcolor}

\begin{document}

\title{QCD Equation of State at very high temperature: \\
computational strategy, simulations and data analysis}

\author{Matteo Bresciani$^{a}$}
\author{Mattia Dalla Brida$^{b,c}$}
\author{Leonardo Giusti$^{b,c}$}
\author{Michele Pepe$^{c}$}
\affiliation{
$^a$ School of Mathematics and Hamilton Mathematics Institute, Trinity College, Dublin, Ireland \\
$^b$ Dipartimento di Fisica, Universit\`a di Milano-Bicocca, Piazza della Scienza 3, I-20126 Milano, Italy\\
$^c$ INFN, Sezione di Milano-Bicocca, Piazza della Scienza 3, I-20126 Milano, Italy}

\begin{abstract}
We present a detailed account of the theoretical progress and the computational strategy 
that led to the non-perturbative determination of the QCD Equation of State at temperatures 
up to the electroweak scale reported in [Phys. Rev. Lett. 134, 201904 (2025)].
The two key ingredients that make such a calculation feasible with controlled uncertainties are: 
(i) the definition of lines of constant physics through the running of a non-perturbatively defined 
finite-volume coupling across a wide range of energy scales, and (ii) the use of shifted boundary 
conditions which allow a direct determination of the entropy density thus without the need 
for a zero-temperature subtraction.
Considering the case of QCD with $N_f =3$ massless flavours in the temperature interval between 3 GeV
and 165 GeV, we describe the numerical strategy based on integrating in the bare 
coupling and quark mass, the perturbative improvement of lattice observables, the optimization of
numerical simulations, and the continuum extrapolation. 
Extensive consistency checks, including finite volume and topology freezing effects, 
confirm the robustness of the method. 
The final results have a relative accuracy of about $1\%$ or better, 
and the errors are dominated by the statistical fluctuations of the Monte Carlo ensembles. 
We also compare our non-perturbative results with predictions from standard 
and hard thermal loop perturbation theory showing that at the level of $\%$-precision contributions 
beyond those known, including non-perturbative ones due to ultrasoft modes, are relevant up to the 
highest temperatures explored. 
The methodological framework is general and readily applicable to QCD with four and five massive 
quark flavours and to other thermal observables, paving the way for systematic non-perturbative 
studies of thermal QCD at very high temperatures.
\end{abstract}

\maketitle


\tableofcontents

\section{Introduction}
\label{sec:Introduction}
The study of the thermodynamics of QCD is central to our understanding of 
strongly interacting matter under extreme conditions, with direct implications for 
cosmology~\cite{Saikawa:2018rcs,Saikawa:2020swg} and heavy-ion physics~\cite{ALICE:2022hor,Gyulassy:2004zy}. 
A key challenge is the reliable determination of equilibrium properties from first principles
across energy scales where perturbation theory is either ineffective or insufficiently accurate. 
Lattice QCD has proven to be the most powerful tool in this domain with full control on the statistical and 
on the systematic effects. 
However, traditional numerical strategies for computing the Equation of State (EoS) by Monte Carlo simulations
face severe limitations when extended to temperatures higher than 1 GeV or so. 
In particular, ultraviolet subtractions and uncertainties in the setting of the bare parameters to 
determine the lines of constant physics,
complicate the realization of this computation and the extraction of results holding in the continuum limit.

To overcome these obstacles, we have developed novel methodologies. 
They combine non-perturbative renormalization techniques based on the running of a renormalized finite-volume gauge 
coupling up to very high energies with shifted boundary conditions in the temporal direction, allowing 
to directly determinate the entropy density which does not need zero-temperature subtractions. 
This framework allows precise simulations at high temperatures while keeping the computational cost 
under control and providing a simple and robust method to extrapolate the results
to the continuum limit at a given physical temperature $T$. 
The approach is general and can be systematically applied to QCD with an arbitrary number of quark flavours, 
including heavy ones, without further theoretical adjustments.

In a recent Letter~\cite{Bresciani:2025vxw}, we reported the first non-perturbative determination of the 
QCD EoS up to the electroweak scale.
The aim of this paper is to present in details the theoretical progress and the computational strategy 
that led to this result.
Building on our earlier determination of the EoS in the SU$(3)$ Yang-Mills theory~\cite{Giusti:2016iqr}, 
we provide here the methodological foundations, numerical techniques, and consistency checks. 
Beyond this immediate application, these tools are of broad relevance to future lattice studies of QCD
thermodynamics, including the extension to theories with four and five quark flavours and to the
determination of other thermal observables, such as transport coefficients and screening masses. 
The tools and techniques presented in this work thus contribute to a broader effort to characterize 
the QCD plasma across all relevant energy scales. 

The paper is organized as follows.
In Section~\ref{sec:Thermal QCD and EoS in the continuum} we first introduce our finite temperature and shifted
boundary conditions setup in the continuum, and present the main formulas for the determination of the EoS.
In Section~\ref{sec:Thermal QCD and EoS on the lattice} we discuss the lattice discretization and the 
renormalization of the lattice theory.
Sections~\ref{sec:Numerical computation} and~\ref{sec:Entropy density} are dedicated to the numerical 
computation of the entropy density of QCD from the lattice, and its extrapolation to the continuum limit.
The full EoS of QCD is finally reported in Section~\ref{sec:Equation of State} where it is also compared with 
the results from the literature at lower temperatures and with the predictions of perturbation theory.
Section~\ref{sec:Conclusions} contains our conclusions and outlook.
The main text of the paper is complemented by an extensive apparatus of Appendices where all the technical 
details of the new strategy proposed here are discussed.

\section{Thermal QCD and the Equation of State in the continuum}
\label{sec:Thermal QCD and EoS in the continuum}
We formulate QCD at finite temperature in the presence of shifted boundary conditions along the compact 
direction~\cite{Giusti:2010bb,Giusti:2011kt,Giusti:2012yj,DallaBrida:2020gux}. 
For the gauge field $A_\mu(x)$, belonging to the algebra of the gauge group SU$(3)$, we have
\begin{equation}
    A_\mu(x_0+L_0, \bs{x}) = A_\mu(x_0, \bs{x}-L_0\vxi )\,,
\label{eq:shBCs_A}
\end{equation}
where $L_0$ is the length of the compact direction and the spatial vector $\vxi=(\xi_1,\xi_2,\xi_3)$ specifies the shift.
We consider $N_f=3$ flavours of quarks in the fundamental representation of the gauge group. 
The related fields are $\psi,\psibar$ and, in the compact direction, they satisfy
\begin{equation}
\begin{aligned}
    \psi(x_0+L_0, \bs{x}) &= -\psi(x_0, \bs{x}-L_0\vxi )\,, \\
    \psibar(x_0+L_0, \bs{x}) &= -\psibar(x_0, \bs{x}-L_0\vxi )\,.
\end{aligned}
\label{eq:shBCs_psi}
\end{equation}
In Euclidean spacetime, this setup describes thermal QCD in a moving reference frame characterized by the boost parameter $\vxi$. 
The free-energy density $f_\vxi$ and the partition function $\cZ_\vxi$ are defined as follows,
\begin{equation}
    f_\vxi = -\frac{1}{L_0 L^3}\ln\cZ_\vxi\,, \quad \cZ_\vxi = \int DAD\psibar D\psi\, e^{-S_{QCD}}\,,
\end{equation}
where $L$ is the size of the three spatial directions, and the thermodynamic limit $L\to\infty$ is understood.
The QCD action $S_{QCD}$ and the conventions for the continuum theory are specified in Appendix B of Ref.~\cite{DallaBrida:2020gux}.
The temperature of the system is $T=1/(L_0\sqrt{1+\vxi^2})$ and it can be varied by changing $L_0$ or the shift $\vxi$.
Since the pressure is $p=-f_\vxi$, the entropy density $s=dp/dT$ can be written as~\cite{Giusti:2012yj}
\begin{equation}
    \frac{s}{T^3} = \frac{1+\vxi^2}{\xi_k}\frac{1}{T^4}\frac{\partial f_\vxi}{\partial \xi_k}\,,
    \label{eq:entropy_cont}
\end{equation}
when $L_0$ is kept fixed. The derivative in the shift removes the additive, ultraviolet power-like divergence in the free-energy. 
The pressure $p(T)$ can be obtained by integrating the entropy with respect to the temperature, while the energy density $e(T)$ follows
from the thermodynamic relation $Ts=e+p$.

\section{Thermal QCD and the Equation of State on the lattice}
\label{sec:Thermal QCD and EoS on the lattice}
We discretize QCD on a 4-dimensional hypercubic lattice of size $L_0/a\times (L/a)^3$, where $a$ is the lattice spacing.
The action of the lattice theory $S_{QCD} = S_G + S_F$ is the sum of the pure gauge action $S_G$ and the fermionic one $S_F$.
\footnote{We use the same notation for lattice and continuum quantities, as any ambiguity is resolved from the context.}
For the former we consider the Wilson plaquette discretization~\cite{Wilson:1974sk},
while the latter is the non-perturbatively $O(a)$-improved Wilson-Dirac action for $N_f=3$ degenerate quark 
flavours~\cite{Wilson:1975hf,Sheikholeslami:1985ij}.
We refer to Appendix~\ref{app:Lattice QCD action} for the explicit expressions.
The lattice fields satisfy shifted boundary conditions along the compact direction analogous to the ones 
in Eqs.~\eqref{eq:shBCs_A} and~\eqref{eq:shBCs_psi}, with the field $A_\mu(x)$ replaced by the link field $U_\mu(x) \in$ SU$(3)$.
Along the spatial directions they satisfy periodic boundary conditions.
On the lattice, the values of the shift should satisfy the constraints $L_0\xi_k/a\in{\mathbb Z}$ 
and $-L/2\leq L_0\xi_k< L/2$, $k=1,2,3$.

\subsection{Renormalization and lines of constant physics}
\begin{table}
	\centering
	\begin{tabular}{|c|c|c|}
		\hline
		$T$  & $T$ (GeV) &  $\bar{g}^2_{\rm SF}(\mu=1/L_0)$\\
		\hline
		$T_0$ &  164.6(5.6) &  1.01636 \\
		$T_1$ &  82.3(2.8)  &  1.11000 \\
		$T_2$ &  51.4(1.7)  &  1.18446 \\
		$T_3$ &  32.8(1.0)  &  1.26569 \\
		$T_4$ &  20.63(63)  &  1.3627  \\   
		$T_5$ &  12.77(37)  &  1.4808  \\
		$T_6$ &  8.03(22)   &  1.6173  \\
		$T_7$ &  4.91(13)   &  1.7943  \\
		$T_8$ &  3.040(78)  &  2.0120  \\
		\hline
	\end{tabular}
	\caption{Second column: physical temperatures considered in this work. 
            Third column: values of the Schr\"odinger functional coupling in $\Nf=3$ QCD at the renormalization scale $\mu=1/L_0$.}
  \label{tab:T0T8GeV}	
\end{table}
Following Ref.~\cite{DallaBrida:2021ddx}, we determine the lines of constant physics at a given temperature $T$ by 
matching the value of the Schr\"odinger functional (SF) coupling $\bar{g}^2_{\rm SF}$ at finite lattice 
spacing to its value in the continuum at a scale $\mu=1/L_0$,
\begin{equation}
    \bar{g}^2_{\rm SF}(g_0^2, a\mu) = \bar{g}^2_{\rm SF}(\mu)\,, \quad a\mu\ll 1\,.
    \label{eq:SFrencond}
\end{equation}
The right-hand side of this equation, i.e. the non-perturbative running of the renormalized coupling
$\bar{g}^2_{\rm SF}(\mu)$, is known precisely in the continuum for QCD with $\Nf=3$ flavours of 
massless degenerate quarks~\cite{Campos:2018ahf,DallaBrida:2018rfy,Bruno:2017gxd,DallaBrida:2016uha}.
The Eq.~\eqref{eq:SFrencond} then fixes the dependence of the bare coupling 
$g_0$ on the lattice spacing, for values of $a$ at which the scale $\mu$ and, therefore, the temperature $T$ 
can be easily accommodated. As a consequence, each temperature can be simulated at several
lattice resolutions, and the continuum limit can be taken with confidence.
Table~\ref{tab:T0T8GeV} reports the 9 values of temperature we considered
in this work, from $3$ GeV up to $165$ GeV. The corresponding values of the SF coupling are also shown.
The critical mass $m_{\rm cr}$ at a given $g_0^2$ and $L_0/a$ is then defined by requiring the PCAC mass to vanish 
in the SF setup. 
We refer to Appendix B of Ref.~\cite{DallaBrida:2021ddx} for the technical details.

\subsection{Entropy density on the lattice}
On the lattice, at fixed bare parameters $L_0/a$ and $g_0^2$, we write the entropy density in Eq.~\eqref{eq:entropy_cont} as follows, 
\begin{equation}
    \frac{s}{T^3} = \frac{1+\vxi^2}{\xi_k}\frac{1}{T^4}
    \frac{\Delta f_\vxi}{\Delta\xi_k}\,,
    \label{eq:entropy_lattice}
\end{equation}
where 
\begin{equation}
    \frac{\Delta f_\vxi}{\Delta\xi_k} = 
    \frac{L_0}{4a}\left(f_{\vxi+\frac{2a}{L_0}\hat{k}} - f_{\vxi-\frac{2a}{L_0}\hat{k}}\right)
    \label{eq:f_diff_discrete_shift}
\end{equation}
is the two-point symmetric discretization of the derivative of the free-energy density with respect to 
the $k$-th component of the shift (see also Appendix~\ref{app:Choice of the shift parameter} for further details).
It is convenient from the computational viewpoint to decompose, at fixed bare parameters, 
the discrete derivative of the free-energy density into two contributions,
\begin{equation}
    \frac{\Delta f_\vxi}{\Delta\xi_k} = 
    \frac{\Delta f_\vxi^\infty}{\Delta\xi_k} + 
    \frac{\Delta (f_\vxi - f_\vxi^\infty)}{\Delta\xi_k}\,,
    \label{eq:split}
  \end{equation}
where $f_\vxi^\infty$ is the free-energy density of QCD with infinitely heavy quarks, i.e. in the static limit of QCD.
We rewrite the first term as follows,
\begin{multline}
    \frac{\Delta f_\vxi^\infty}{\Delta\xi_k} = 
    \frac{\Delta f^{(0),\infty}}{\Delta\xi_k}
    + g_0^2\, \frac{\Delta f^{(1),\infty}}{\Delta\xi_k} \\
    - \int_0^{g_0^2} du \left(\frac{1}{u}\left.\frac{\Delta\corrno{\overline{S_G}}_\vxi^\infty}{\Delta\xi_k}\right|_{g_0^2=u}
    +\frac{\Delta f^{(1),\infty}}{\Delta\xi_k}\right)\,,
    \label{eq:Df_gauge}
\end{multline}
where $f^{(0),\infty}$ and $f^{(1),\infty}$ represent, respectively, the tree-level and one-loop coefficients of 
the expansion in lattice perturbation theory of the free-energy density (see 
Appendix~\ref{sec:Free-energy density in lattice perturbation theory}) computed at infinite bare quark masses.
The discrete derivatives in the shift are defined analogously to Eq.~\eqref{eq:f_diff_discrete_shift}, and
\begin{equation}
	\corr{\overline{S_G}}^{\infty}_\vxi = \frac{a^4}{L_0L^3}\corr{S_G}^{\infty}_\vxi
	\label{eq:corrSG_volavg}
\end{equation}
is the expectation value of the pure gauge action density in the static quark limit and in the shifted setup.
The second contribution to Eq.~\eqref{eq:split} can be rewritten as
\begin{multline}
   \frac{\Delta (f_\vxi - f_\vxi^\infty)}{\Delta\xi_k} =  
   - \frac{\Delta}{\Delta\xi_k} \int_0^\infty dm_q\frac{\partial f_\vxi^{m_q}}{\partial m_q} \\
   = -\int_0^\infty dm_q \frac{\Delta\corrno{\psibar\psi}_\vxi^{m_q}}{\Delta\xi_k}\,,
   \label{eq:Df_quark}
\end{multline}
where $m_q = m_0 - m_{\rm cr}(g_0^2, L_0)$ is the bare subtracted quark mass and $\corrno{\psibar\psi}_\vxi^{m_q}$ is
the expectation value of the scalar density at a given subtracted quark mass.
Again, the discrete derivative with respect to the shift is defined similarly to Eq.~\eqref{eq:f_diff_discrete_shift}.

\section{Numerical computation}
\label{sec:Numerical computation}
In the following we discuss the numerical evaluation on the lattice of
the two contributions in Eqs.~\eqref{eq:Df_gauge} and~\eqref{eq:Df_quark}.
At each given temperature we have considered four resolutions $L_0/a=4,6,8,10$ of the compact direction, 
while the three spatial directions are of the same size $L/a = 144$: the aspect ratio $LT$ of our lattices 
thus ranges between $10$ and $25$.
The choice of the shift is to some extent arbitrary, and can be leveraged for a good compromise between discretization effects
and relative accuracy of the entropy density, as we discuss in Appendix~\ref{app:Choice of the shift parameter}.
The chosen value in this work is $\vxi=(1,0,0)$.
At a given temperature and $L_0/a$, the values of the inverse bare coupling, of the critical hopping parameter and of the 
non-perturbative $O(a)$-improvement coefficient can be found in Table 4 of Ref.~\cite{DallaBrida:2021ddx}.

\subsection{Numerical determination of \mtht{$\static$}}
\label{ssec:Static contribution}
We computed the contribution $\static$ by estimating the integral in the bare coupling 
of Eq.~\eqref{eq:Df_gauge} using an optimized combination of numerical quadratures.
A similar integration has been employed in Ref.~\cite{Giusti:2015daa}, where it was shown that the integrand 
function is smooth in $g_0^2$ in the interval of interest, and that any systematic effects 
arising from the numerical quadratures are negligible with respect to the statistical accuracy of the integral.
At a given $L_0/a$ and for each value of $g_0^2$ prescribed by the chosen quadrature rules, we have measured $\corr{\overline{S_G}}^\infty_\vxi$ 
in simulations of the pure SU$(3)$ Yang-Mills theory in the shifted setup, 
at the values $\vxi=(1\pm2a/L_0, 0, 0)$ required by the computation of the discrete derivative with respect to the shift.
The integral in the bare coupling is then obtained from the linear combination of these results, with weights that are
uniquely determined by the quadrature rules and the integration intervals.

\begin{figure}[h]
    \centering
    \includegraphics[width=\columnwidth]{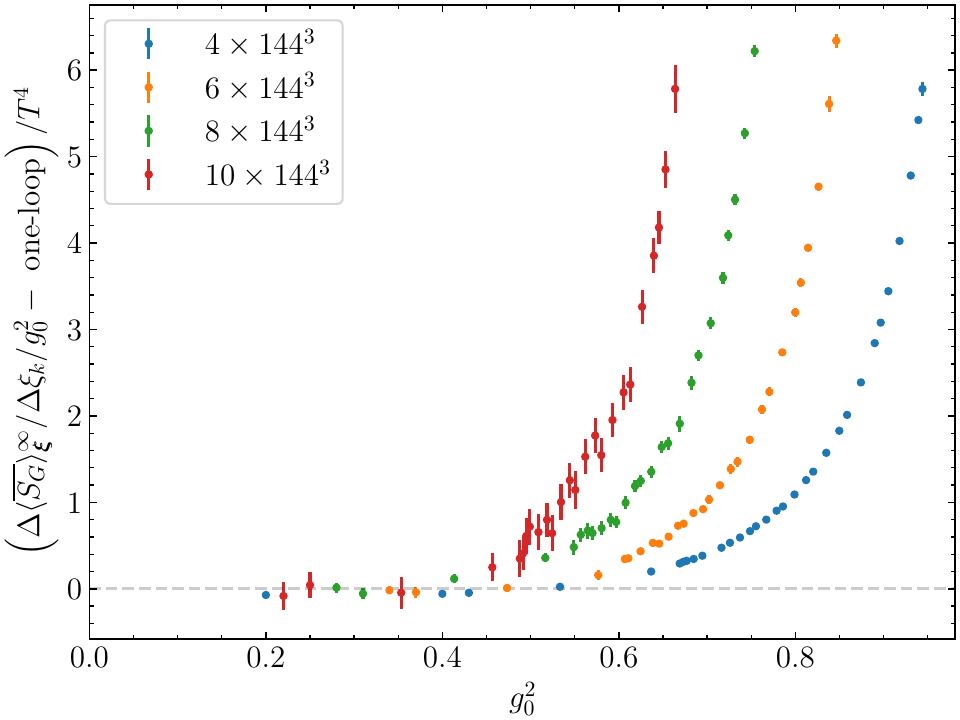}
    \caption{
    Plot of the integrand function in Eq.~\eqref{eq:Df_gauge} as a function of the bare coupling $g_0^2$.
    Points have been shifted horizontally by $0.03\times(L_0/a-4)$ for better readability.
    }
    \label{fig:g02_integrand_06x144}
\end{figure}

At fixed $L_0/a$, the integration in the bare coupling has to be performed in the 
intervals $g_0^2\in[0, g_0^2|_{T_i}]$, $i = 0, 1,. . ., 8$, where $g_0^2|_{T_i}$ is the bare coupling 
squared at the given temperature $T_i$ whose corresponding inverse values $6/g_0^2$ are reported in Table~\ref{tab:bare_results_DfDxi}. 
We have computed first the integral for the value of $g_0^2$ associated to the temperature $T_1$
by splitting the integration in three domains as follows,
\begin{equation}
    \int_0^{g_0^2|_{T_1}} \hspace{-1.2em}  du\, (\bullet)=
    \int_0^{6/15} \hspace{-1.em}  du\, (\bullet)
    + \int_{6/15}^{6/9} \hspace{-0.75em}  du\, (\bullet)
    + \int_{6/9}^{g_0^2|_{T_1}} \hspace{-1.2em}  du\, (\bullet),
    \label{eq:split_integral_g02}
\end{equation}
where the dot stands for the integrand function of Eq.~\eqref{eq:Df_gauge}.
For $L_0/a=4,6$ we have one further temperature, $T_0$, whose integral is computed as in Eq.~\eqref{eq:split_integral_g02}
by replacing $g_0^2|_{T_1}$ with $g_0^2|_{T_0}$.
At the lower temperatures ($i>1$), the integral is then obtained by adding to 
the result at temperature $T_{i-1}$ the integral in the domain $g_0^2\in[g_0^2|_{T_{i-1}},\,g_0^2|_{T_i}]$,
\begin{equation}
    \int_0^{g_0^2|_{T_i}}du\,(\bullet) = \int_0^{g_0^2|_{T_{i-1}}}du\,(\bullet) + \int_{g_0^2|_{T_{i-1}}}^{g_0^2|_{T_i}}du\,(\bullet)\,.
    \label{eq:from_Ti_to_Tj}
\end{equation}
A summary of the integration scheme can be found in Table~\ref{tab:tab_g02_integral}.
\footnote{Note that the integration starting from $g^2_0=0$ is carried out up to values of the bare coupling well below the critical point where 
the first order transition of the SU($3$) pure gauge theory occurs, see e.g. Ref.~\cite{Giusti:2025fxu}.}
The first column reports the integration domains, and the second column the respective quadrature rules and number of integration points for 
the different values of $L_0/a$ and temperatures.
A showcase of the numerical results for $\frac{\Delta\corr{\overline{S_G}}_\vxi^\infty}{\Delta\xi_k}$ is given in Table~\ref{tab:DSGDxi}
of Appendix~\ref{app:Further numerical results} for $L_0/a=6$ together with the values of inverse bare coupling $6/g_0^2$.
The integrand function of Eq.~\eqref{eq:Df_gauge} is shown in Figure~\ref{fig:g02_integrand_06x144} for the four lattice 
resolutions $L_0/a=4,6,8,10$.
\begin{table}
    \centering
    \begin{tabular}{|c|lc|}
        \hline
        Interval & \multicolumn{2}{c|}{Quadrature} \\
        \hline
        \multirow{2}{*}{$0\leq g_0^2\leq 6/15$} & 3 (Simpson) & $L_0/a=4$ \\
                                                & 2 (trapezoid) & $L_0/a=6,8,10$ \\
        \hline
        $6/15 \leq g_0^2 \leq 6/9$              & 3 (Gauss-Legendre) &    \\
        \hline
        \multirow{2}{*}{$6/9\leq g_0^2\leq g_0^2|_{T_0}$} & 3 (Gauss-Legendre)    & $L_0/a=4$ \\
                                                          & 1 (midpoint) & $L_0/a=6$ \\
        \hline
        $6/9 \leq g_0^2 \leq g_0^2|_{T_1}$                & 3 (Gauss-Legendre)  & \\
        \hline
        \multirow{2}{*}{$g_0^2|_{T_{i-1}}\leq g_0^2\leq g_0^2|_{T_i}$} & 3 (Gauss-Legendre)    & $1<i<7$ \\
                                                                      & 5 (Gauss-Legendre)    & $i=7,8$ \\
        \hline
    \end{tabular}
    \caption{Summary of the integration scheme for the computation of the integral in $g_0^2$ appearing in Eq.~\eqref{eq:Df_gauge}.}
    \label{tab:tab_g02_integral}
\end{table}

The pure gauge ensembles for this computation have been generated by Monte Carlo simulations 
where the basic sweep is a combination of heatbath and over-relaxation~\cite{Adler:1987ce} updates of the link variables, 
using the Cabibbo–Marinari scheme~\cite{Cabibbo:1982zn,Creutz:1979dw,Creutz:1980zw}.
We collected about 6000 measurements of $S_G$ per quadrature point (about 3000 per shift) for $L_0/a=4,6$ 
and about 40000 measurements per quadrature point for the two finest resolutions $L_0/a=8,10$.
The final results for $\static$ at fixed lattice spacing are reported in Table II of Ref.~\cite{Bresciani:2025vxw}
and they have relative errors of a few permille for $L_0/a=4,6$, about $0.5\%$ for $L_0/a=8$ and about $1.5\%$ for $L_0/a=10$.

\subsection{Numerical determination of \mtht{$\dynamic$}}
\label{ssec:Dynamic contribution}
We now describe the numerical strategy for the computation of the integral
in the bare quark mass that defines the contribution $\dynamic$ to the QCD entropy density through Eq.~\eqref{eq:Df_quark}.
We split the integral in three parts,
\begin{multline}
    \int_0^\infty dm_q \frac{\Delta\corrno{\psibar\psi}_\vxi^{m_q}}{\Delta\xi_k}
    = T\int_0^{5}d\mt_q \, (\bullet) \\
    + T\int_{5}^{\mt}d\mt_q\, (\bullet) + T\int_{\mt}^\infty d\mt_q\, (\bullet)\,,
    \label{eq:integral_mq_split}
\end{multline}
where $\mt_q=m_q/T$ and $\mt = 35$ for $L_0/a=4$ or $\mt=20$ for $L_0/a=6,8,10$.
We have chosen a 10-point Gauss quadrature for the first domain and a 7-point Gauss quadrature for the second domain
(6-point Gauss quadrature for $L_0/a=6$, $6/g_0^2=8.5403$). 
We have integrated the third domain with a 3-point Gauss quadrature, after a change of integration variable to the hopping parameter
\begin{equation}
    \kappa = \frac{1}{2(aT\mt_q+a\mcr+4)}
    \label{eq:changeofvariable}
\end{equation}
which makes this integration interval compact.
As explained in Appendix~\ref{sec:Systematics from the numerical quadratures}, this optimized integration scheme 
guarantees that any systematic error introduced by the numerical quadratures is negligible with respect to the statistical 
accuracy of the non-perturbative results.
\begin{figure*}[t]
   \begin{center}
   \begin{minipage}{\columnwidth}
   \includegraphics[width=\textwidth]{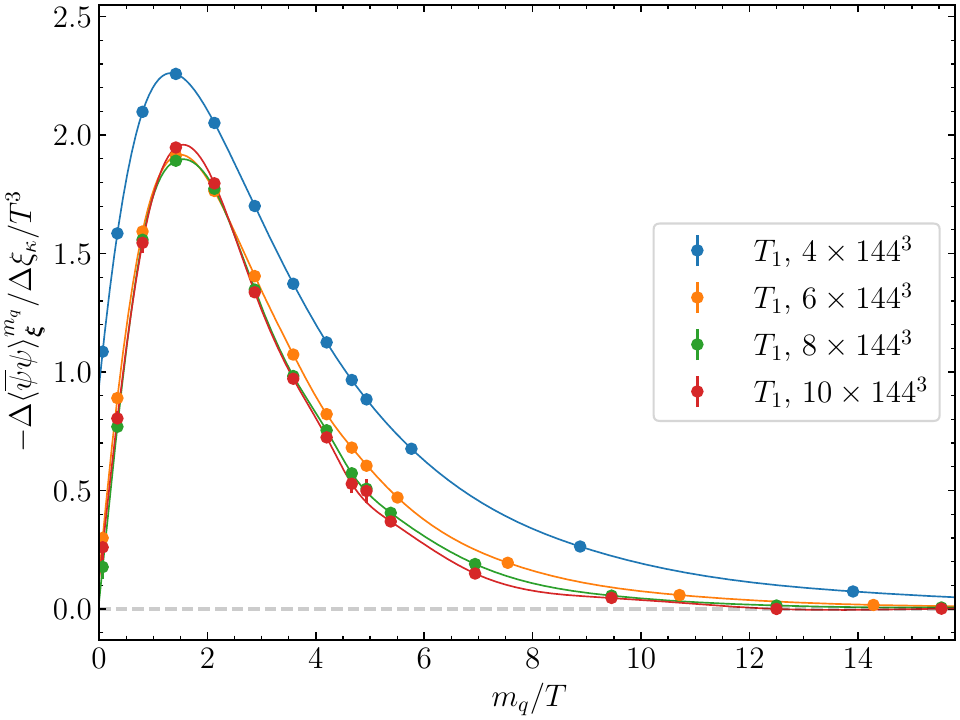}
   \end{minipage}
   \begin{minipage}{\columnwidth}
   \includegraphics[width=\textwidth]{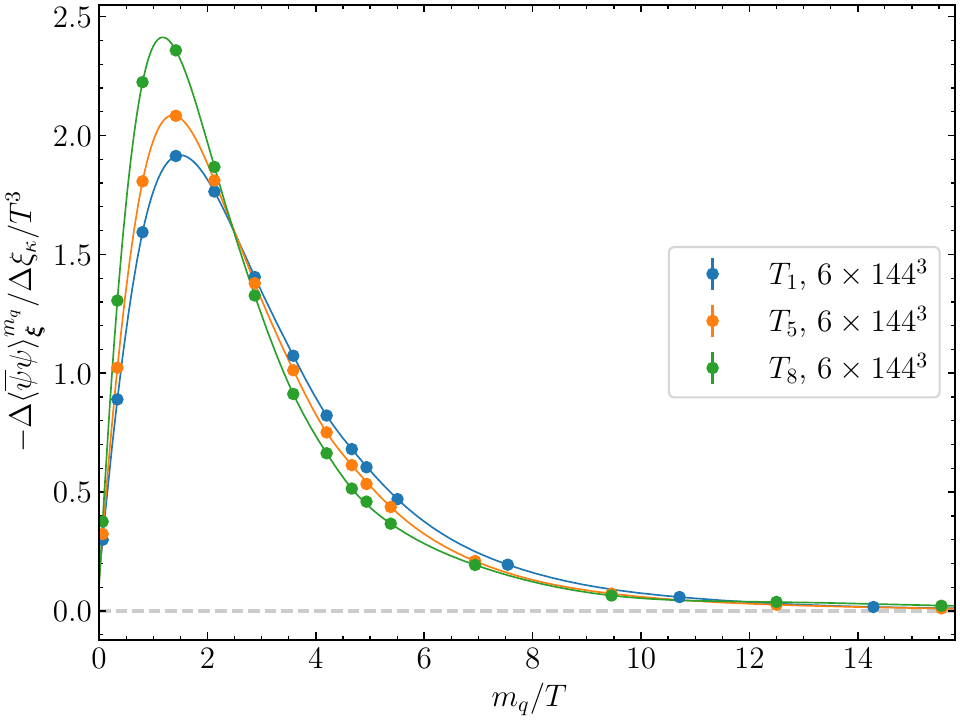}
   \end{minipage}
   \end{center}
   \caption{
   Left: plot of the integrand function in Eq.~\eqref{eq:Df_quark}
   computed at the bare parameters of temperature $T_1$ and at the resolutions $L_0/a=4,6,8,10$, 
   as a function of $m_q/T$. Points have been interpolated with a cubic spline to guide the eye.
   In most cases, errors are smaller than the markers.
   Right: the same integrand function is shown at the resolution $L_0/a=6$ and at three temperatures.
   }
   \label{fig:mq_integrand_06x144}
\end{figure*}
The non-perturbative integrand function is represented in Figure~\ref{fig:mq_integrand_06x144} for different values of the lattice spacing
(left panel) and different temperatures (right panel); we report in Table~\ref{tab:DSdDxi} the numerical results of $\frac{\Delta\corrno{\psibar\psi}_\vxi^{m_q}}{\Delta\xi_k}$ at some representative values of the bare parameters.

The non-perturbative results for $\frac{\Delta(f_\vxi-f_\vxi^\infty)}{\Delta\xi_k}$
at given $L_0/a$ and $g_0^2$ are collected in Table II of Ref.~\cite{Bresciani:2025vxw}. 
The numerical determination of each of them requires to simulate QCD at the 20 values 
of bare quark mass prescribed by the nodes of the Gauss quadratures, for the two shifts $\vxi=(1\pm 2a/L_0,0,0)$ 
giving a total of 40 independent simulations. Taking advantage of the peculiarities of the quantity of interest,
we managed to highly optimize the numerical approach in order to minimize the computational effort.
We summarize these aspects in the following, and refer to dedicated Appendices for more details.

\subsubsection{Simulating at large quark masses}
The values of bare subtracted quark mass in our QCD simulations range
between the chiral limit $m_q=0$ and the static quark limit $m_q\to\infty$.
The heavier the quarks, the less they contribute to the dynamics:
this allows us to treat the fermionic forces in the molecular dynamics of the hybrid Monte Carlo (HMC) with coarser 
integration schemes as we increase the bare quark mass, with small impact on the acceptance rate. 
By scrutinizing in detail the magnitude of the fermionic forces and the 
spectral gap of the Dirac operator for increasing values of the bare quark mass, we selected four optimal 
algorithms for the molecular dynamics which maximize the acceptance rate (always above $90\%$) 
of the HMC while minimizing the computational cost.
The description of the chosen algorithms, of the tuning procedure and of the generation of 
our QCD ensembles can be found in Appendix~\ref{app:Details on lattice QCD simulations}.

\subsubsection{Variance reduction}
\label{ssec:Variance reduction}
The primary quantity for the numerical determination of $\dynamic$ is the expectation value of the 
scalar density, $\corr{\psibar\psi}_\vxi^{m_q}$, to be computed in QCD simulations with increasing values of the bare quark mass.
To profit from translational invariance, at fixed gauge configuration we have employed U$(1)$ random sources
for the computation of the trace of the quark propagator.
Following Ref.~\cite{Giusti:2019kff} we have introduced an improved estimator of the latter, obtained by subtracting
its leading non-trivial order in the hopping parameter expansion.
This significantly reduces the contribution to the variance due to the random sources with respect to the 
naive estimator, with an overall gain up to a factor $2.5$ in the statistical error of $\dynamic$
at negligible additional computational effort.
The details of this procedure are reported in Appendix~\ref{app:Scalar density variance reduction}.

\subsubsection{Optimization of the statistics}
At fixed parameters $L_0/a$ and $g_0^2$, the uncertainty of the contribution $\dynamic$ is propagated 
from the one of $\corr{\psibar\psi}_\vxi^{m_q}$, which is measured in independent numerical simulations at the 
different values of the bare quark mass prescribed by the Gauss quadratures.
The different magnitude of the Gauss weights and the dependence of the variance of the scalar density estimator with the 
bare quark mass (see Appendix~\ref{app:Scalar density variance reduction}) make the Gauss points contribute differently 
to the final error.
The computational effort changes with the bare quark mass too, and it is possible to 
tune the statistics for $\corr{\psibar\psi}_\vxi^{m_q}$ at each Gauss point such that the total 
computational cost is minimized given a target precision for $\dynamic$.
This optimization has been performed for a target relative error of $0.5\%$ for $L_0/a=4,6,8$, and $1\%$ for $L_0/a=10$.
Further details on this procedure can be found in Appendix~\ref{sec:Optimization_of_the_statistics},
where the optimized number of measurements is reported.
This optimization led to a final gain up to a factor $2$ in the computational time compared to the case
where, at fixed target accuracy, the same statistics is considered for all the Gauss points.

\section{Entropy density}
\label{sec:Entropy density}

\begin{table}[t]
   \centering
   \begin{tabular}{|c|cc|cc|cc|}
   \hline
         & & & & & & \\[-0.25cm]
   $L_0/a$ & $6/g_0^2$ & $\frac{\Delta f_\vxi}{\Delta\xi_k}\times10^4$ &
             $6/g_0^2$ & $\frac{\Delta f_\vxi}{\Delta\xi_k}\times10^4$ &
             $6/g_0^2$ & $\frac{\Delta f_\vxi}{\Delta\xi_k}\times10^4$  \\ [0.125cm]
   \hline
   \hline
      & \multicolumn{2}{c|}{$T_0$} & \multicolumn{2}{c|}{$T_1$} & \multicolumn{2}{c|}{$T_2$} \\
   \hline
    4 & 8.7325 &  149.05(4) & 8.3033 &  148.29(5) & 7.9794 &  147.55(6) \\
    6 & 8.9950 &   22.07(6) & 8.5403 &   21.92(4) & 8.2170 &   21.96(4) \\
    8 &   -    &     -      & 8.7325 &  6.448(26) & 8.4044 &  6.443(23) \\
   10 &   -    &     -      & 8.8727 &  2.596(21) & 8.5534 &  2.581(22) \\
   \hline
      & \multicolumn{2}{c|}{$T_3$} & \multicolumn{2}{c|}{$T_4$} & \multicolumn{2}{c|}{$T_5$} \\
   \hline
    4 & 7.6713 &  146.87(6) & 7.3534 &  146.07(6) & 7.0250 &  144.93(6) \\
    6 & 7.9091 &   21.80(5) & 7.5909 &   21.72(5) & 7.2618 &   21.58(5) \\
    8 & 8.0929 &  6.388(25) & 7.7723 &  6.389(26) & 7.4424 &  6.403(29) \\
   10 & 8.2485 &  2.567(22) & 7.9322 &  2.597(23) & 7.6042 &  2.550(26) \\
   \hline
      & \multicolumn{2}{c|}{$T_6$} & \multicolumn{2}{c|}{$T_7$} & \multicolumn{2}{c|}{$T_8$} \\
   \hline
    4 & 6.7079 &  143.80(7) & 6.3719 &  142.32(9) &   -    &     -      \\    
    6 & 6.9433 &   21.50(5) & 6.6050 &   21.38(6) & 6.2735 &   21.14(7) \\    
    8 & 7.1254 &  6.326(23) & 6.7915 &    6.32(4) & 6.4680 &    6.30(4) \\    
   10 & 7.2855 &  2.577(26) & 6.9453 &  2.562(23) & 6.6096 &  2.511(26) \\    
   \hline                                                  
   \end{tabular}
   \caption{Results at fixed bare parameters of the derivative in the shift of the free-energy density.}
   \label{tab:bare_results_DfDxi}
\end{table}
The results discussed in this Section correspond to those reported in Ref.~\cite{Bresciani:2025vxw}, where the EoS of
$N_f = 3$ QCD was determined non-perturbatively. Here, we provide the full details of the continuum extrapolation and of the fit strategy
adopted.

At fixed $L_0/a$ and $g_0^2$, the derivative of the free-energy density with respect to the shift, $\frac{\Delta f_\vxi}{\Delta\xi_k}$, 
is obtained by summing the two contributions $\static$ and $\dynamic$ determined as described in Section~\ref{sec:Numerical computation}; 
the final results are reported in Table~\ref{tab:bare_results_DfDxi}.
The entropy density then follows from Eq.~\eqref{eq:entropy_lattice}.
In propagating the errors from the primary observables to the entropy density, we have accounted for correlations using the
Gamma-method~\cite{Wolff:2003sm}, as implemented in Refs.~\cite{Joswig:2022qfe,Ramos:2018vgu}.
We have also included the systematic uncertainty coming from the definition of the lines of constant 
physics~\cite{DallaBrida:2018rfy,DallaBrida:2021ddx}, even though its contribution is negligible within the final accuracy.

\subsection{Finite volume effects}
\label{ssec:Finite volume effects}
On general grounds, finite volume effects in QCD are exponentially suppressed as $e^{-M_{\rm gap}L}$, 
where $M_{\rm gap}$ is the mass gap of the theory and $L$ is the linear spatial size of the system.
In the high-temperature regime under investigation, $M_{\rm gap}$ corresponds to the mass of the lightest screening state, 
which scales proportionally to $T$ with a coefficient close to unity~\cite{Giusti:2012yj,DallaBrida:2021ddx,Laine:2009dhh}.
Since in our simulations $10\lesssim LT\lesssim 25$, finite size effects are expected to be negligible compared to the 
the statistical accuracy of our numerical results.

In order to validate this expectation, in Ref.~\cite{Bresciani:2025vxw} we presented a comparison of non-perturbative results from lattices
with spatial sizes $L/a=144$ and $L/a=288$ at some selected bare parameters. Instead of computing the entropy density from the derivative of
the free-energy density, we used the one-point function of the energy-momentum tensor~\cite{Giusti:2012yj}. 
This provides an alternative discretization of the entropy density, 
\begin{equation}
    \frac{s}{T^3} = - \frac{1+\vxi^2}{\xi_k} \frac{1}{T^4} \langle T_{0k}^{R,\{6\}}\rangle_{\vec\xi}\,,
    \label{eq:s_T0k6}
\end{equation}
equivalent to Eq.~\eqref{eq:entropy_lattice} up to cutoff effects.
In Eq.~\eqref{eq:s_T0k6} the sextet components of the renormalized energy-momentum tensor,
\begin{equation}
    T_{\mu\nu}^{R,\{6\}} = Z_G^{\{6\}}(g_0^2)\,T_{\mu\nu}^{G,\{6\}} + Z_F^{\{6\}}(g_0^2)\,T_{\mu\nu}^{F,\{6\}}\,,
    \label{eq:T6R}
\end{equation}
are defined as in Section 3 of Ref.~\cite{DallaBrida:2020gux}.
At present, the entropy density cannot be determined non-perturbatively from
Eq.~\eqref{eq:s_T0k6} because the renormalization constants $Z_G^{\{6\}}$ and 
$Z_F^{\{6\}}$ are known only at one-loop order in bare lattice perturbation theory.
However, since the latter are insensitive to finite volume effects, the matrix 
elements $\corr{T_{\mu\nu}^{G,\{6\}}}_\vxi$ and $\corr{T_{\mu\nu}^{F,\{6\}}}_\vxi$ can 
be used as a proxy to study the finite volume effects affecting the entropy density. 
Their lattice determination is significantly less demanding than the computation of the entropy via Eq.~\eqref{eq:entropy_lattice}.

We computed the bare matrix elements on $6\times 144^3$ and $6\times 288^3$ lattices, with a statistical precision
such that the relative error on $s/T^3$ as in Eq.~\eqref{eq:s_T0k6} is comparable to that obtained
from the results in Table~\ref{tab:bare_results_DfDxi}.
The renormalization constants $Z_G^{\{6\}}$ and $Z_F^{\{6\}}$, appearing in Eq.~\eqref{eq:T6R}, have been estimated using perturbation 
theory at one-loop order~\cite{DallaBrida:2020gux}.
We carried out this study at the temperatures $T_1$, $T_8$, and 
no finite volume effects were observed in the matrix elements (see Table III in Ref.~\cite{Bresciani:2025vxw}).
This makes us confident that, for $10\lesssim LT\lesssim 25$, they are negligible on the entropy density at all the temperatures considered in this work.

\subsection{Restricting to zero topological sector}
\label{ssec:Restricting to zero topology sector}
In the interval of temperatures that we have investigated, $3\,{\rm GeV}\lesssim T \lesssim 165\,{\rm GeV}$, the trivial topological sector
of the QCD phase space gives by far the dominant contribution to the path integral. 
At asymptotically high temperatures, the instanton analysis predicts the topological susceptibility to be suppressed as 
$\sim T^{-b}m^3$, $b\sim8$, for QCD with three light degenerate quark flavours of mass $m$.
The analogous prediction for the pure SU$(3)$ gauge theory has been explicitly verified non-perturbatively~\cite{Giusti:2018cmp}.
Similarly, lattice QCD calculations tend to confirm the scaling with $T$ predicted in the semi-classical analysis, even though
the systematics introduced by dynamical fermions is still difficult to control~\cite{Borsanyi:2016ksw}.
For all practical purposes, we can compute the entropy density by restricting
our main observables, $\corr{\overline{S_G}}_\vxi^\infty$ and $\corr{\psibar\psi}_\vxi^{m_q}$, to the trivial topological sector.
The systematics introduced by neglecting non-zero topological sectors is much below the statistical accuracy of our numerical results.

In our pure gauge ensembles we have however observed some topological activity for bare couplings between $g_0^2|_{T_7}$ and
$g_0^2|_{T_8}$ for $L_0/a=6,8,10$ and for $6/g_0^2\lesssim6.5$ when $L_0/a=4$. In the QCD simulations, non-vanishing topological
sectors occurred only in $L_0/a=4,6$ ensembles at temperatures $T_7, T_8$ and for values of the hopping parameter $\kappa\lesssim0.10$,
where the topological susceptibility is expected to be less suppressed. We have computed the expectation values of the Wilson plaquette
action and of the scalar density restricting the integration in various topological sectors, and no difference can be observed within
statistical errors. This result is in agreement with other studies at lower temperatures~\cite{Bazavov:2017dsy}.
So as to be the most conservative, we added a systematic error to $\corr{\overline{S_G}}_\vxi^\infty$ 
for data between $T_7$ and $T_8$ up to a total relative error of $2\%$, 
in order to safely take into account any effect from the observed topological fluctuations.

\subsection{Continuum limit}
\label{ssec:Continuum limit}
Improved definitions of observables reduce the systematic effects of lattice artifacts when performing the extrapolation 
to the continuum limit of results obtained at finite lattice spacing. 
For the entropy density we have considered the following improved definition,
\begin{equation}
   s\left(\frac{a}{L_0}, g_0^2\right) \to s\left(\frac{a}{L_0}, g_0^2\right) \,
         \frac{\dfrac{\partial}{\partial\xi_k} \left[f_0 + g^2\,f_1\right] }
              {\dfrac{\Delta }{\Delta\xi_k}\left[ f^{(0)} + g^2\, f^{(1)}\right] }\,,
   \label{eq:pt_improvement}
\end{equation}
where $g=\bar{g}_{\rm SF}(1/L_0)$, $f_0$ and $f_1$ are the tree-level and one-loop perturbative coefficients of the free-energy density 
in the continuum while $f^{(0)}$, $f^{(1)}$ are the corresponding coefficients computed in lattice perturbation 
theory~\cite{DallaBrida:2020gux}, summarized in Appendix~\ref{sec:Free-energy density in lattice perturbation theory}. 
Table~\ref{tab:entropy_lpt} contains the results at fixed lattice spacing and in the continuum limit relevant for the 
improvement at one-loop order.
Using Eq.~\eqref{eq:pt_improvement}, discretization effects at tree-level and $O(g^2)$ are subtracted from our 
observable to all orders in the lattice spacing. 
\begin{table}
    \centering
    \begin{tabular}{|c|ccccc|}
    \hline
    $a/L_0$ & $1/4$ & $1/6$ & $1/8$ & $1/10$ & $0$ \\
    \hline
    $\frac{1}{T^4}\frac{\Delta f^{(0)}}{\Delta\xi_k}$ & 16.561 &   12.065 &   11.036 &   10.736 &   10.418 \\
    [0.15cm]
    $\frac{1}{T^4}\frac{\Delta f^{(1)}}{\Delta\xi_k}$ & $-2.086$ &   $-1.163$ &   $-0.885$ &   $-0.808$ &  $-0.750$ \\
    \hline
    \end{tabular}
    \caption{Coefficients for improvement of the lattice entropy density at one-loop order. 
    The last column corresponds to the continuum values appearing in Eq.~(\ref{eq:pt_improvement}).} 
    \label{tab:entropy_lpt}
\end{table}

\begin{table*}[t]
\begin{center}
    \begin{tabular}{|c|c|c|c||c||c|c|c|}
    \hline
               & \texttt{id0} & \texttt{id1} & \texttt{id2} & \texttt{id3} (best fit) & \texttt{id4} & \texttt{id5} & \texttt{id6} \\
    \hline
       $L_0/a$ & $4,6,8,10$ & $4,6,8,10$ &   $6,8,10$ &   $6,8,10$ &   $6,8,10$ &   $6,8,10$ &   $6,8,10$ \\
          syst &            &            &            &   $a^3g^3$ &   $a^3g^3$ &   $a^4g^3$ &   $a^3g^4$ \\
    \hline
         $c_0$ & 20.057(22) &   20.00(8) &   20.10(7) &   20.13(8) &  20.14(11) &   20.11(7) &   20.15(6) \\
         $c_1$ & 20.031(25) &   19.97(9) &   20.01(6) &   20.05(8) &  20.06(11) &   20.01(7) &   20.05(6) \\
         $c_2$ & 19.996(28) &  19.93(10) &   20.04(7) &   20.05(9) &  20.06(11) &   20.04(7) &   20.07(7) \\
         $c_3$ &   19.97(3) &  19.89(11) &   19.90(7) &   19.90(9) &  19.91(11) &   19.90(8) &   19.92(8) \\
         $c_4$ &   19.94(3) &  19.86(12) &   19.88(8) &  19.93(10) &  19.93(11) &   19.89(9) &   19.92(9) \\
         $c_5$ &   19.88(4) &  19.79(13) &   19.81(9) &  19.87(11) &  19.87(11) &  19.82(10) &  19.86(10) \\
         $c_6$ &   19.84(4) &  19.73(15) &  19.73(10) &  19.75(12) &  19.75(12) &  19.73(10) &  19.73(11) \\
         $c_7$ &   19.80(5) &  19.67(18) &  19.69(12) &  19.74(15) &  19.73(15) &  19.70(13) &  19.70(14) \\
         $c_8$ &   19.63(7) &  19.48(22) &  19.53(14) &  19.58(17) &  19.56(20) &  19.55(15) &  19.51(17) \\
    \hline
      $d_{23}$ &     5.5(3) &      10(7) &    6.9(1.9) &       6(4) &      3(16) &    6.7(2.4) &            \\
      $d_{24}$ &            &            &            &            &      2(12) &            &    5.1(2.5) \\
      $d_{33}$ &            &    -15(22) &            &            &            &            &            \\
    \hline
    $\chi^2/\chi^2_{\rm exp}$ &       1.06 &       1.12 &       0.74 &       0.82 &       0.85 &       0.76 &       0.80 \\
    \hline
    \end{tabular}
    \caption{Summary of the results of the fits that we considered for the extrapolation of $s/T^3$ to the continuum limit.
    Each column of the table corresponds to a fit, labelled for convenience by the \texttt{id}-number in the first row.
    The second row tells which lattice spacings have been included in the fit.
    The third row indicates if a systematic error arising from the indicated cutoff effects has been added in quadrature to 
    the statistical error of the numerical results, see main text for details.
    The remaining rows report the fitted parameters and the value of $\chi^2/\chi^2_{\rm exp}$.
    }
   \label{tab:fits_comparison}
\end{center}
\end{table*}
We have extrapolated to the continuum limit the one-loop improved lattice results for the entropy density by a global fit 
of the data to all temperatures. 
We have parametrized the temperature dependence of the discretization effects in terms of polynomials of a renormalized coupling, and
the natural choice is the non-perturbative SF coupling used to define the lines of constant physics.
Since the lattice theory is $O(a)$-improved, and since at finite temperature odd powers in the coupling are generally present, 
after the improvement at one-loop order the leading discretization 
effects of the lattice results are expected to be of order $O(a^2g^3)$. 
Thus, we considered the following general fit function,
\begin{multline}
    s(T_i, a/L_0)/T_i^3 = c_i
    + \left(\frac{a}{L_0}\right)^2 \left( d_{23}\,g_i^3 + d_{24}\,g_i^4 \right) \\
    + \left(\frac{a}{L_0}\right)^3 \left( d_{33}\,g_i^3 + d_{34}\,g_i^4 \right)\,,
    \label{eq:fitfunc}
\end{multline}
where $i=0,...,8$ and $g_i=\bar{g}_{\rm SF}(\sqrt{2}T_i)$.
The fit parameters $c_i$ are the continuum results for $s/T^3$ at the different temperatures $T_0,...,T_8$, while
the $d_{ij}$ parametrize the discretization effects.
We explored a variety of fits to assess the cutoff effects and ensure that systematic uncertainties
from the extrapolation were well under control.
We report on the more representative fits, which are also summarized in Table~\ref{tab:fits_comparison}, 
each labelled with an \texttt{id} number.

We first considered fits without $O(g^4)$ terms by enforcing $d_{24}=d_{34}=0$.
We fitted data with $L_0/a=4,6,8,10$ by setting either $d_{33}=0$ (fit \texttt{id0}) or $d_{33}\neq0$ (fit \texttt{id1}).
While both fits have $\chi^2/\chi^2_{\rm exp}\approx 1$, with $\chi_{\rm exp}$ defined as in Ref.~\cite{Bruno:2022mfy}, 
fit \texttt{id0} gives continuum values $c_i$ with errors $3$-$4$ times smaller.
Although compatible within the larger errors of fit \texttt{id1}, the extrapolated central values of the first fit are systematically higher.

A quadratic fit of the lattice artifacts excluding the data at $L_0/a = 4$ yields 
$\chi^2/\chi^2_{\rm exp}=0.74$ (fit \texttt{id2}). 
This fit provides estimates of the continuum values that are in good agreement with fit \texttt{id1}, 
both in terms of central values and error size.
This suggests that the data at $L_0/a=4$ are likely affected by discretization effects of higher order than $a^2$. 
Thus, we used the data at the coarsest lattice spacing only to estimate 
the size of the $O(a^3)$ contributions and included these as a systematic error for the $L_0/a=6,8,10$ data.

More specifically, we used the value of $d_{33}$ obtained from the fit \texttt{id1} as an estimate of the systematic 
uncertainty and we added it in quadrature to the statistical errors, $\sigma(T_i, a/L_0)$, of $s/T^3$ at given temperature and lattice spacing:
\begin{equation}
   \sigma^2\left(T_i, \frac{a}{L_0}\right) \to \sigma^2\left(T_i, \frac{a}{L_0}\right) + \left[\left(\frac{a}{L_0}\right)^3 d_{33} \, g_i^3\right]^2\,.
   \label{eq:inflation}
\end{equation}
The final best fit is the one considering data with $L_0/a>4$, $d_{24}=d_{33}=d_{34}=0$ in the fit ansatz Eq.~\eqref{eq:fitfunc}, 
and errors as in Eq.~\eqref{eq:inflation} for the definition of the weights in the $\chi^2$-function minimized by the fit.
This fit is labelled \texttt{id3} in Table~\ref{tab:fits_comparison}. The continuum results for the normalized entropy 
density have a relative error of $0.5$-$1.0\%$, and their covariance matrix can be found in Appendix~\ref{app:Further numerical results}.
Figure~\ref{fig:clim} shows the related continuum extrapolation.
\begin{figure}[t]
    \centering
    \includegraphics[width=\columnwidth]{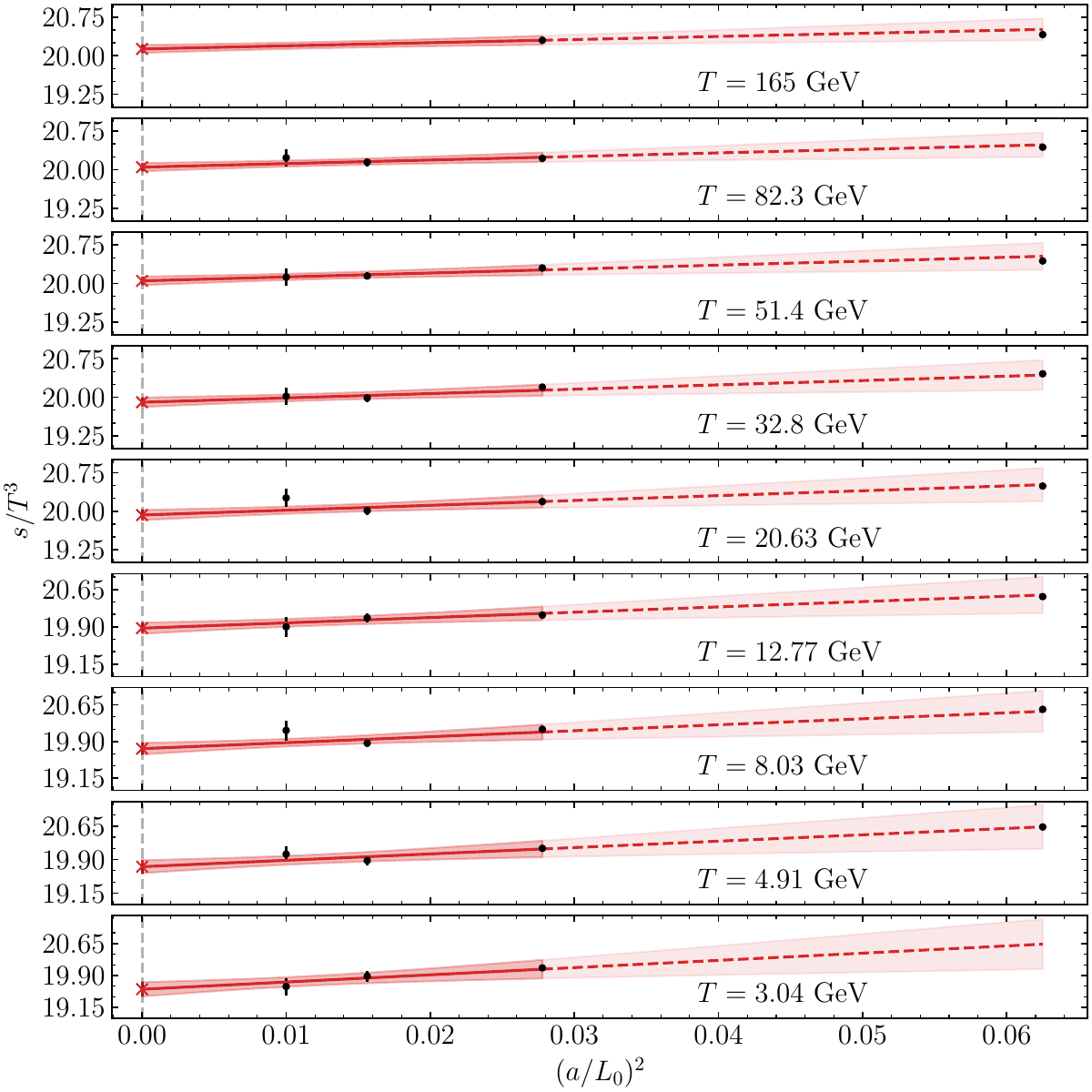}
    \caption{Black dots are the values of the one-loop improved normalized entropy density as a function of $(a/L_0)^2$ 
    at the temperatures $T_0, T_1, ..., T_8$.
    The red band is our best extrapolation to the continuum limit (fit \texttt{id3} in Table~\ref{tab:fits_comparison}).
    Red crosses are the continuum extrapolated values for $s/T^3$.
    The horizontal axis is common to all the subplots.
    }
    \label{fig:clim}
\end{figure}

We performed several checks to further corroborate the robustness of our best fit.
We first repeated the best fit letting $d_{24}\neq0$ in Eq.~\eqref{eq:fitfunc}.
The continuum limits of this fit \texttt{id4} are perfectly compatible to those of the best fit, 
meaning that the parameterization of the cutoff effects is not sensitive (within errors) to the 
inclusion of higher powers of the renormalized coupling.
We also repeated the whole analysis considering subleading cutoff effects of 
$O(a^4)$ instead of $O(a^3)$ in the fit ansatz and in Eq.~\eqref{eq:inflation}.
The resulting $c_i$ (fit \texttt{id5}) are stable with respect to the best fit, with errors that are $10$-$20\%$ smaller. 
Similar conclusions hold when, in the cutoff effects, the leading power of the coupling is chosen to be $O(g^4)$ instead of $O(g^3)$
in both the fit ansatz and in the determination of the systematic error (fit \texttt{id6}).
\begin{figure*}[t]
   \begin{center}
   \begin{minipage}{\columnwidth}
   \includegraphics[width=\textwidth]{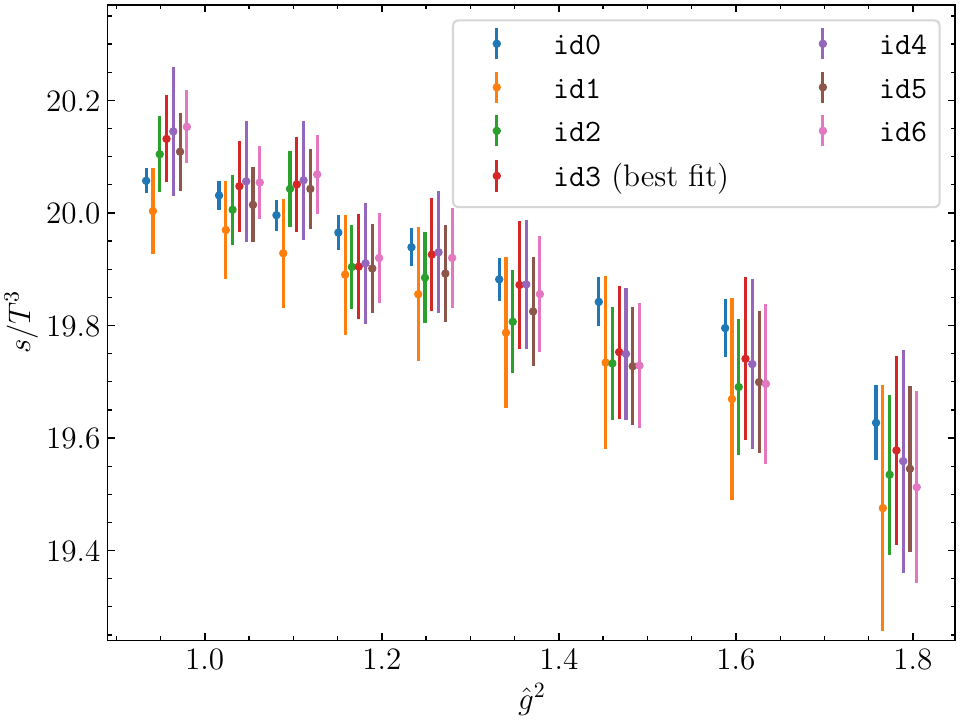}
   \end{minipage}
   \begin{minipage}{\columnwidth}
   \includegraphics[width=\textwidth]{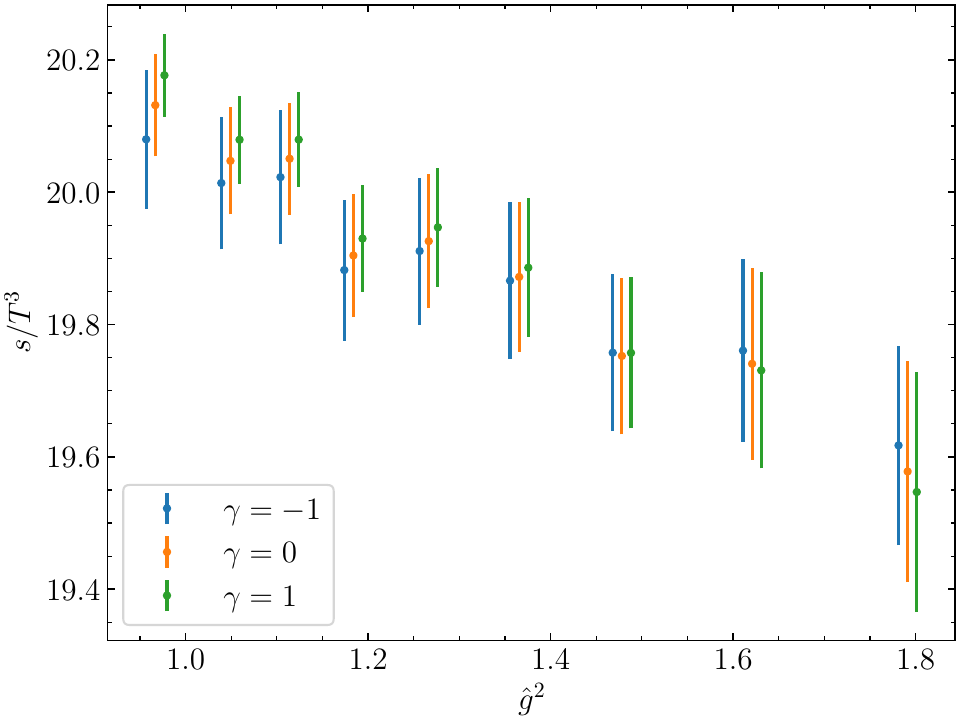}
   \end{minipage}
   \end{center}
    \caption{Left: comparison among the results of different continuum limits.
    Points have been shifted horizontally by $0.023\times(n/3-1)$ where $n=0,1,...,6$ for fits \texttt{id0}, \texttt{id1}, ..., \texttt{id6}.
    Right: effect of logarithmic corrections on the continuum extrapolated values of the best fit.
    For better readability points have been shifted horizontally by $0.01\times\gamma$, with $\gamma$
    defined in Eq.~\eqref{eq:log_corr}.
    }
    \label{fig:clim_comparison} 
\end{figure*}

The continuum results from all these fits are actually well compatible with each other as shown in the left panel of 
Figure~\ref{fig:clim_comparison}.
In the latter, data have been plotted against the renormalized coupling squared $\hat g^2$, defined as the five-loop 
$\overline{\rm MS}$ coupling~\cite{Baikov:2016tgj} at the renormalization scale $\mu=2\pi T$, whose leading order expression is 
\begin{equation}\label{eq:gmu}
  \frac{1}{\hat g^2(\mu)} = \frac{9}{8\pi^2} \ln  \frac{\mu}{\Lambda_{\MSbar}} +\ldots \, , \quad \mu=2\pi T\,,
\end{equation}
where $\Lambda_{\MSbar} = 341$~MeV is taken from Ref.~\cite{Bruno:2017gxd}.
For our purposes, this is only a convenient function of $T$ that we choose to study the temperature dependence 
of our non-perturbative results.
Further details on the computation of $\hat g$ are reported in Appendix~\ref{app:Strong coupling at five-loop in MSbar scheme}.

We finally checked the impact of logarithmic corrections to the leading discretization effects, of $O(a^2g^3)$, 
on the best fit \texttt{id3} using the modified fit ansatz~\cite{Husung:2019ytz,Husung:2021mfl,Husung:2022kvi}
\begin{equation}
    s\left(T_i, a/L_0\right)/T_i^3 = c_i + d_{23} \left[\bar g_{\rm SF}^2(\pi/a)\right]^\gamma
    \left(\frac{a}{L_0}\right)^2 g_i^3\,,
    \label{eq:log_corr}
\end{equation}
where $\bar g_{\rm SF}(\pi/a)$ is the renormalized SF coupling evaluated at the cutoff scale.
The continuum results of $s/T^3$ change by less than one standard deviation with respect
to the  best fit, $\gamma=0$, when the effective anomalous dimension is varied in the interval $\gamma\in[-1,1]$. 
The comparison for three selected values of $\gamma$ is shown in the right panel of Figure~\ref{fig:clim_comparison}.

\section{Equation of State}
\label{sec:Equation of State}
The main non-perturbative results are the values of the normalized entropy density $s/T^3$ of continuum QCD reported in 
Table~\ref{tab:fits_comparison} with the fit label \texttt{id3}, corresponding to the 9 temperatures from 3 GeV to 165 GeV 
of Table~\ref{tab:T0T8GeV}.
These values are shown as black dots in Figure~\ref{fig:bests_comp} as a function of the strong coupling squared
$\hat g^2$ defined above.
These results can be fitted to obtain an analytic expression for the temperature dependence of the entropy density, and thus
for the other thermodynamic functions in the EoS, namely the pressure and the energy density.

\subsection{Temperature dependence of the entropy density}
\label{ssec:Temperature dependence of the entropy density}
\begin{figure}[ht]
    \centering
    \includegraphics[width=\columnwidth]{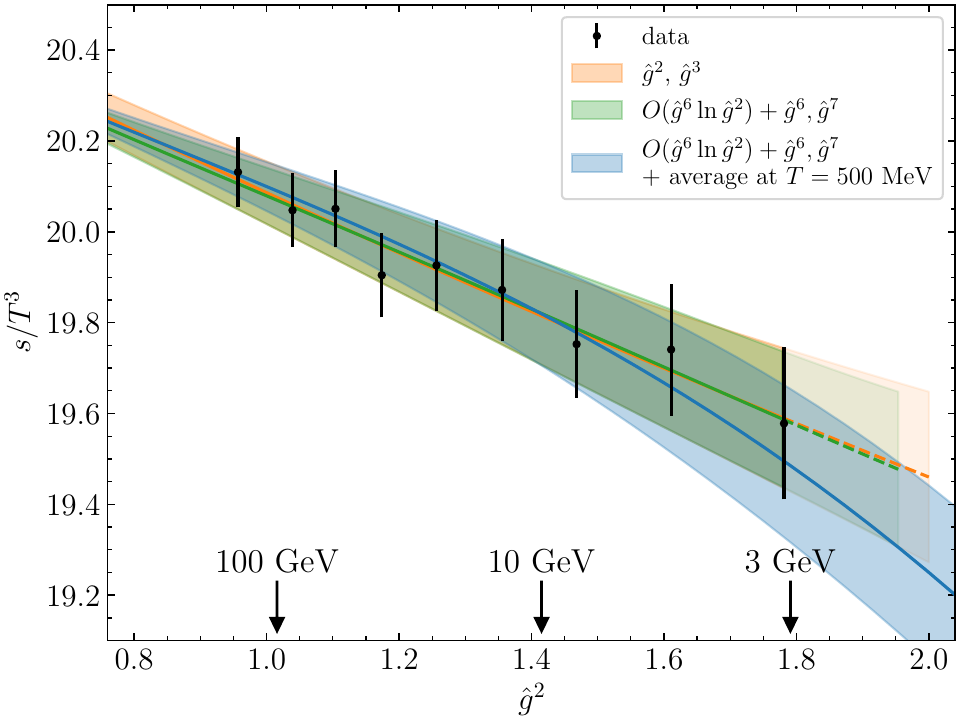}
    \caption{Comparison of the three parametrizations of the temperature dependence of $s/T^3$, in the 
    temperature interval covered by the non-perturbative data.}
    \label{fig:bests_comp}
\end{figure}

We have parametrized the temperature dependence of the entropy density in terms of a polynomial in the renormalized coupling $\hat g$,
\begin{equation}
    \frac{s}{T^3} = \frac{32\pi^2}{45}\sum_k s_k\left(\frac{\hat{g}}{2\pi}\right)^k\,.
    \label{eq:s_mikko}
\end{equation}
This convenient choice allows us to more easily compare our non-perturbative results with the predictions of perturbation theory.
This study has been carried out in Ref.~\cite{Bresciani:2025vxw}: we report in Table~\ref{tab:fits_ghat2} the results 
for the fits which are relevant for the discussion in the following.
The first fit has been performed by enforcing the Stefan-Boltzmann result at infinite temperature, $s_0^{\rm SB}=2.969$, and taking 
the coefficients $s_2$, $s_3$ as fit parameters.
This fit is shown in Figure~\ref{fig:bests_comp} as the orange shadowed band.

In the second fit we have enforced all the known coefficients in perturbation theory up to $O(\hat g^6\ln \hat g^2)$
~\cite{Kajantie:2002wa,Shuryak:1977ut,Chin:1978gj,Kapusta:1979fh,Toimela:1982hv,Arnold:1994ps,Arnold:1994eb,Zhai:1995ac} 
so as to constrain the parametrization in the asymptotically high temperature regime.
We have fitted to the data the unknown term at $O(\hat g^6)$ and one further term of $O(\hat g^7)$ to take into account 
any higher order effects in the expansion in $\hat g$.
The green band in Figure~\ref{fig:bests_comp} shows this fit.
The fitted coefficients actually include the non-perturbative contribution of the ultrasoft modes~\cite{Linde:1980ts,Gross:1980br}
and a not yet known perturbative term.
The determination of the missing perturbative coefficient at $O(\hat g^6)$ (see Ref.~\cite{Navarrete:2024ruu} for recent progresses) 
would allow us to disentangle the ultrasoft effects and estimate the relative magnitude.

\begin{figure*}[ht]
\begin{center}
\begin{minipage}{\columnwidth}
\includegraphics[width=\textwidth]{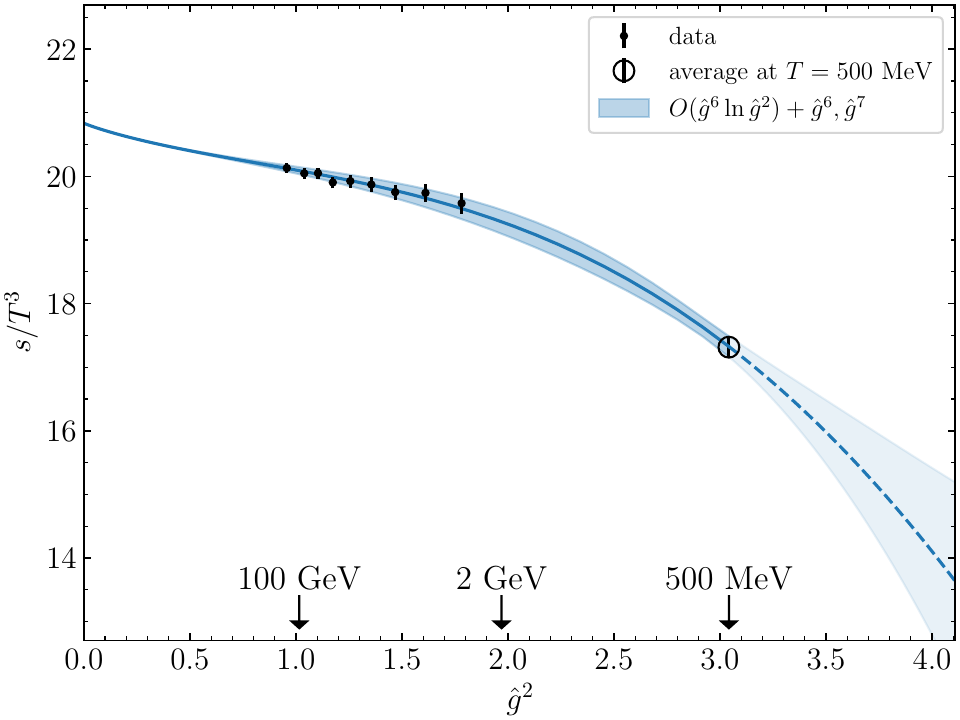}
\end{minipage}
\begin{minipage}{\columnwidth}
\includegraphics[width=\textwidth]{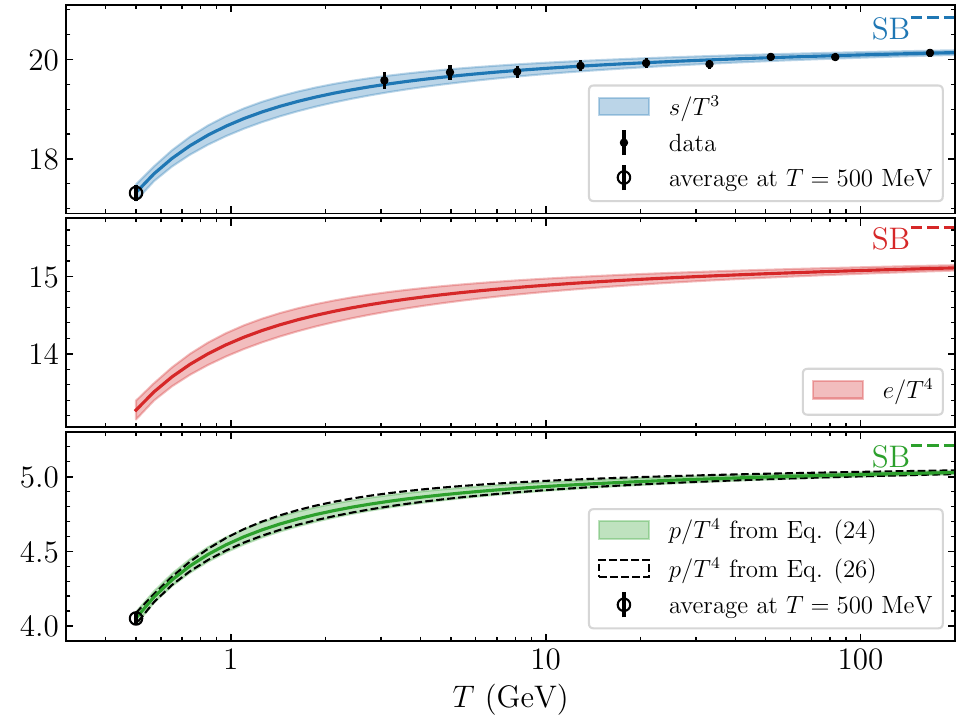}
\end{minipage}
\caption{Left: Normalized entropy density $s/T^3$ in the continuum limit as a function of $\hat g^2$.
         The blue band is the best parametrization for $T\geq500$ MeV, and includes the point at $T=500$ MeV obtained 
         from Refs.~\cite{Borsanyi:2013bia,HotQCD:2014kol,Bazavov:2017dsy} (open circle marker).
         Right: EoS as a function of the temperature for $T\geq 500$ MeV.}
\label{fig:best_sep}
\end{center}
\end{figure*}

The third fit ansatz is the same as the second, but we included in the non-perturbative dataset the value $s/T^3=17.31(16)$ at 
$T=500$ MeV coming from the weighted average of the results from Refs.~\cite{Borsanyi:2013bia,HotQCD:2014kol,Bazavov:2017dsy}, obtained
in $N_f=2+1$ QCD. 
This procedure is consistent as the effect of the quark masses at $T=500$ MeV is several times smaller than the error on the entropy 
density quoted by the two collaborations~\cite{Laine:2006cp}.
This fit is our best parametrization of the entropy density of $N_f=3$ QCD for $T\geq 500$ MeV, and it is represented in 
Figure~\ref{fig:bests_comp} (blue band) and in the left panel of Figure~\ref{fig:best_sep}.
Figure~\ref{fig:bests_comp} shows that all these parametrizations are perfectly compatible in the whole temperature 
range from $3$ GeV to $165$ GeV covered by the non-perturbative results, both in the central values and in the error bands.

\begin{table*}[th]
\centering
\begin{tabular}{|c|c|c|c||c|}
   \hline
   $k$ & $s_k$ (orange fit) & $s_k$ (green fit) & $s_k$ (blue fit) & $p_k$ \\
   \hline
   0 & $2.969$   & $2.969$                             & $2.969$                             & $2.969$  \\
   1 &   $0$     &    $0$                              &    $0$                              &   $0$    \\
   2 & $-5.1(9)$ & $-8.438$                            & $-8.438$                            & $-8.438$ \\
   3 &  $5(5)$   & $55.11$                             & $55.11$                             & $55.11$  \\
   4 &   $0$     & $-40.28 + 101.2\ln\hat g^2$         & $-40.28 + 101.2\ln\hat g^2$         & $-49.77 + 101.2\ln\hat g^2$ \\
   5 &   $0$     & $-1174$                             & $-1174$                             & $-1081$ \\
   6 &   $0$     & $4791 - 1629\ln\hat g^2 -5.1(1.7)\times 10^3$  & $4791 - 1629\ln\hat g^2 -4.0(1.1)\times 10^3$  & $4776 - 1401\ln\hat g^2 -4.0(1.1)\times 10^3$ \\
   7 &   $0$     & $1.3(7)\times 10^4$                            &  $0.7(4)\times 10^4$                           & $0.4(4)\times 10^4$ \\
   \hline
\end{tabular}
\caption{The first three columns contain the results from fitting the entropy density $s/T^3$ to a polynomial in the coupling $\hat g$ 
according to Eq.~\eqref{eq:s_mikko}.
Values with errors are the fit parameters, while the other coefficients are set either to zero or to their perturbative value 
obtained from Ref.~\cite{Kajantie:2002wa} (see main text for the details). 
In the first row, colours refer to Figure~\ref{fig:bests_comp}.
The last column contains the coefficients for the pressure $p/T^4$, parametrized as in Eq.~\eqref{eq:p_mikko}.
}
\label{tab:fits_ghat2}
\end{table*}

\begin{figure*}[ht]
\begin{center}
\begin{minipage}{\columnwidth}
\includegraphics[width=\textwidth]{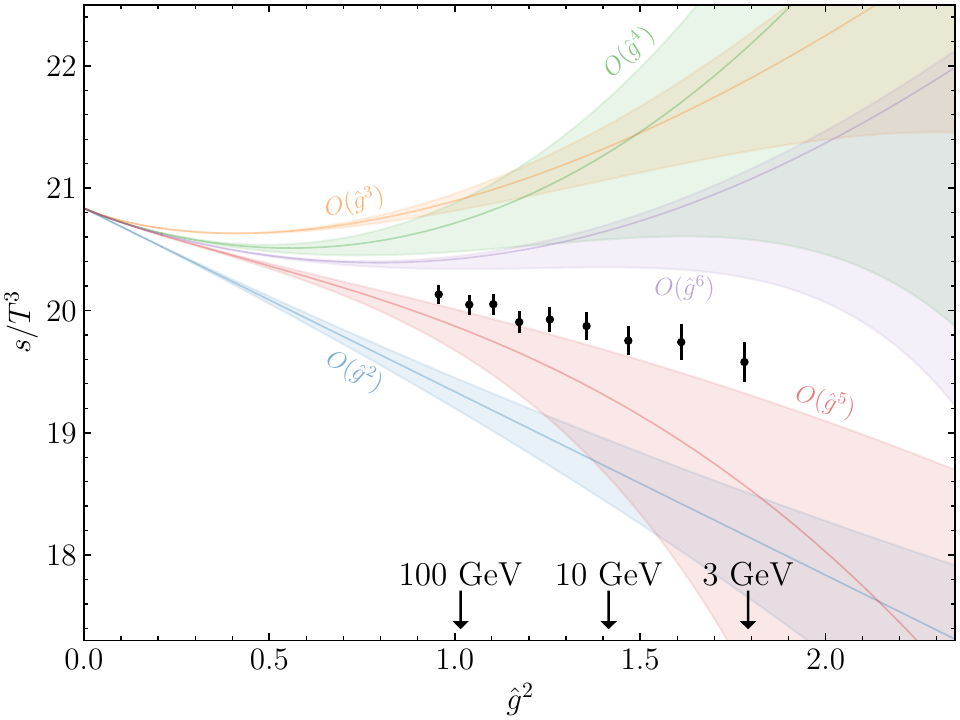}
\end{minipage}
\begin{minipage}{\columnwidth}
\includegraphics[width=\textwidth]{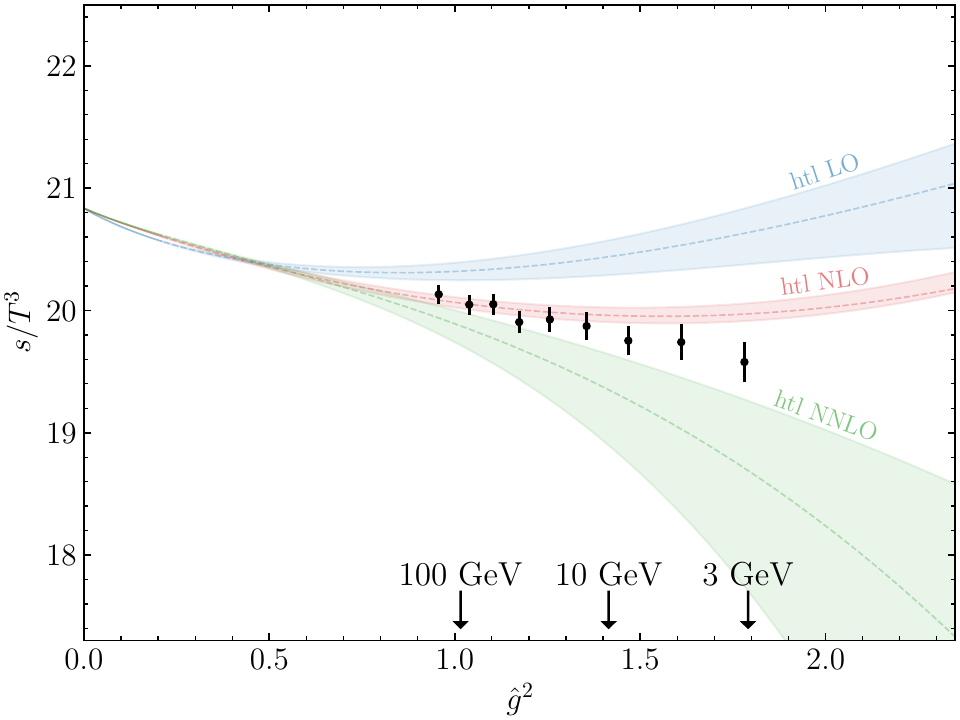}
\end{minipage}
\caption{Black dots are the non-perturbative values of $s/T^3$ plotted against $\hat g^2$.
Some values of the physical temperature are reported for reference.
In the left plot, the curves represent the predictions of perturbation theory
obtained from Ref.~\cite{Kajantie:2002wa}, each including up to the order in $\hat g$ indicated by the label.
In the right plot, the curves represent the hard thermal loop perturbation 
theory~\cite{Andersen:2003zk,Andersen:2010ct,Andersen:2011sf}
at leading order (LO), next-to leading order (NLO) and next-to-next-to leading order (NNLO).
In both plots, error bands are obtained by a variation of the renormalization scale $\mu=2\pi T$ by the factors $0.5$ and $2$.}
\label{fig:comp_pt_htl}
\end{center}
\end{figure*}

\subsection{Comparison with perturbation theory}
\label{ssec:Comparison with perturbation theory}
We now compare our non-perturbative results for the normalized entropy density with the 
analytic results in standard perturbation theory and in hard thermal loop perturbation 
theory~\cite{Andersen:2003zk,Andersen:2010ct,Andersen:2011sf}.

The curves of the perturbative expansion for $s/T^3$ are shown in the left panel of Figure~\ref{fig:comp_pt_htl}, 
each including all the orders up to $O(\hat g^k)$, $k=2,3,...,6$ as indicated by the labels. 
The shadowed bands correspond to varying the renormalization scale by the factors $0.5$ and $2$ with respect to $\mu=2\pi T$. 
We have also propagated the statistical uncertainty on the non-perturbative $\Lambda_\MSbar$ \cite{Bruno:2017gxd},
but the related contribution is completely subdominant. 
The curves at $O(\hat g^k)$ and $O(\hat g^{k+1})$ with $k=2,3,4,5$ in Figure~\ref{fig:comp_pt_htl} are in tension by a few combined standard deviations, 
with the exception of the $k=3$ case which shows some agreement for $T\lesssim 100$ GeV.
Therefore, in the temperature interval under investigation, higher order effects in the perturbative expansion are relevant and the 
convergence is slow. 
This is also signalled by the fact that, in our best parametrization (third column of Table~\ref{tab:fits_ghat2}), 
the contributions beyond the known perturbative expansion at $T\sim 165$ GeV still give about $40\%$ of the interactions.
It is thus difficult to give a reliable estimation of the systematic uncertainty of the perturbative series as the latter 
is far from the asymptotic regime in the temperature range covered by our non-perturbative data 
(black points in Figure~\ref{fig:comp_pt_htl}).

Similar conclusions can be drawn by considering the hard thermal loop (htl) resummation of the perturbative 
expansion~\cite{Andersen:2011sf}, see also Ref.~\cite{Fernandez:2021sgr} for recent developments. 
The right panel of Figure~\ref{fig:comp_pt_htl} compares our non-perturbative results for $s/T^3$ with the htl curves up to 
next-to-next-to leading order (NNLO). Again, the shadowed bands correspond to a variation of the renormalization scale 
$\mu=2\pi T$ by the factors $0.5$ and $2$, and the contribution to the bands width from the statistical uncertainty of $\Lambda_\MSbar$ 
is negligible. 
Even though the expansion appears to converge better in the asymptotically high temperature regime compared to standard perturbation theory, 
the NLO and NNLO curves are still in tension in the interval of temperature covered 
by our data, with a discrepancy from a few combined standard deviations at $T\sim 3$ GeV down to about $1$ combined standard deviation at $T\sim 165$ GeV.
The agreement between the NNLO prediction with the non-perturbative points is at the level of
$\sim1.5$ standard deviations if we combine the statistical 
error of the points with the systematic one due to the renormalization scale variation.

\subsection{Pressure and energy density}
Given the parametrization for the entropy density (see the third column of Table~\ref{tab:fits_ghat2}), 
we can derive the temperature dependence of the pressure $p(T)$ and of the energy density $e(T)$.
Following Ref.~\cite{Kajantie:2002wa}, we parametrize the pressure analogously to the entropy in Eq.~\eqref{eq:s_mikko},
\begin{equation}
    \frac{p}{T^4} = \frac{8\pi^2}{45}\sum_k p_k\left(\frac{\hat{g}}{2\pi}\right)^k\,,
    \label{eq:p_mikko}
\end{equation}
where the coefficients $p_0,...,p_6$ are given in Table~\ref{tab:fits_ghat2}.
The coefficient $p_7$ can be matched to $s_7$ through the relation $s(T) = \frac{d}{dT}p(T)$, 
leading to 
\begin{equation}
    p_7 = s_7 + \frac{45}{16}\,p_5 + 3\,p_3\,.
\end{equation}
Finally, the energy density is determined using the relation $e=Ts-p$.
The three quantities are shown as functions of the temperature in the right panel of Figure~\ref{fig:best_sep}, for $T\geq 500$ MeV.

As a check for this result we also determined the pressure by integrating the entropy density with respect to the temperature,
\begin{equation}
    p(T) = p(500\,{\rm MeV}) + \int_{500\,{\rm MeV}}^T dT' s(T')\,,
\end{equation}
where $p/T^4=4.050(38)$ at $T=500$ MeV is the weighted average of the results for the pressure computed 
in Refs.~\cite{Borsanyi:2013bia,HotQCD:2014kol,Bazavov:2017dsy}.
The comparison of the pressure as a function of the temperature computed following the two strategies described above can be found
in the right panel of Figure~\ref{fig:best_sep}.
The two determinations are perfectly compatible in the entire temperature range.

\section{Conclusions}
\label{sec:Conclusions}
We have presented a comprehensive strategy for the non-perturbative study of QCD thermodynamics at very high temperatures. The
central ingredients are the definition of lines of constant physics through the running of a renormalized finite-volume coupling across a broad range of
energies and the use of shifted boundary conditions to access directly the entropy density. The combination of these elements allowed us to
overcome the limitations of traditional lattice methods, such as the need of zero-temperature subtractions and the difficulty of matching
bare parameters across widely separated energy scales, like high temperatures and the hadronic scale.  

While the main physical results of this work were previously reported in Ref.~\cite{Bresciani:2025vxw}, the present paper provides the full
computational framework and numerical validation that support those findings. We have described in detail how these ideas are implemented in
practice: the integration techniques in the bare quark mass and in the gauge coupling, the perturbative improvement of lattice observables,
the continuum extrapolation strategy and the dedicated checks of systematic effects. Together, these methodological advances provide a
framework that is efficient and robust to investigate the thermal features of QCD at high temperatures.

Although the present study has focused on the case of three massless flavours, the techniques developed here are of broader scope. Our simulations span nine 
temperatures between 3 GeV and 165 GeV, with multiple lattice spacings at each temperature to enable a reliable continuum extrapolation. The
entropy density is computed directly from the derivative of the free-energy density with respect to the shift parameter, and the pressure
and the energy density are obtained via standard thermodynamic relations. The final results for the EoS are accurate within $1\%$ or better
across the entire temperature range, with uncertainties dominated by statistical fluctuations.  

A detailed analysis of discretization effects, of finite volume corrections and of topological contributions confirms the robustness of our
results. We have also performed a comparison with perturbative predictions, including both standard and hard thermal loop
resummations.
Our findings show that, at our level of accuracy, higher-order contributions in $\hat g^2$ beyond those known, including non-perturbative ones 
due to ultrasoft modes, are relevant even at temperatures as high as 165 GeV.
Furthermore, our results are consistent with existing lattice data at lower temperatures, down to 500 MeV, and a smooth interpolation can be drawn 
across the entire range.

The methodology discussed in this paper is broadly applicable and can be extended to QCD with four or five massive quark flavours. 
By disentangling the methodological framework from its first physical application, this work complements earlier results and sets the stage
for a systematic program of non-perturbative studies of QCD in the high-temperature regime, with potential impact on cosmology, heavy-ion
phenomenology, and the theoretical development of thermal field theory.
In particular, the knowledge of the derivative of the 
free-energy in the shift paves the way to the non-perturbative definition of the QCD
energy-momentum tensor, following the strategy proposed in 
Ref.~\cite{DallaBrida:2020gux}. 
Solving this long standing theoretical problem
\cite{Caracciolo:1989pt,Caracciolo:1989bu,Caracciolo:1991vc,Caracciolo:1991cp}
would enable further first-principles investigations of the thermal properties of QCD through correlation functions of the energy-momentum tensor.

\acknowledgements
We acknowledge PRACE for awarding us access to the HPC system MareNostrum4 
at the Barcelona Supercomputing Center (Proposals n. 2018194651 and 2021240051)
where some of the numerical results presented in this paper have been obtained. 
We also thank CINECA for providing us with a very generous access to Leonardo 
during the early phases of operations of the machine and for the computer time 
allocated via the CINECA-INFN, CINECA-Bicocca agreements. 
The R\&D has been carried out on the PC clusters Wilson and Knuth at 
Milano-Bicocca. 
We thank all these institutions for the technical support. 
This work is (partially) supported by ICSC – Centro Nazionale di Ricerca in High
Performance Computing, Big Data and Quantum Computing, funded by European 
Union – NextGenerationEU.

\appendix

\section{Lattice QCD action}
\label{app:Lattice QCD action}
The lattice QCD action, $S_{QCD}=S_G+S_F$, is decomposed in the pure gauge part $S_G$ and in the fermionic one $S_F$.
For the former we employ the Wilson plaquette action~\cite{Wilson:1974sk},
\begin{equation}
    S_G = \frac{6}{g_0^2}\sum_x\sum_{\mu<\nu}\left[ 1 - \frac{1}{3}{\rm Re}\,{\rm tr}\left\{U_{\mu\nu}(x)\right\}\right]\,,
    \label{eq:SG}
\end{equation}
where $g_0$ is the bare coupling, $U_{\mu\nu}$ are the plaquette fields
\begin{equation}
  U_{\mu\nu}(x)=U_\mu(x)\, U_\nu(x+ a \hat{\mu})\, U_\mu(x+ a \hat{\nu})^\dag\, U_\nu(x)^\dag\,,
  \label{eq:plaquette}
\end{equation}
$\hat{\mu},\hat{\nu}$ are unit vectors oriented along the directions $\mu,\nu$ respectively, 
and $U_\mu(x)\in{\rm SU}(3)$ are the link fields.
The fermionic action is
\begin{equation}
    S_F=a^4\sum_x \psibar(x) (D+M_0)\psi(x)
    \label{eq:SF}
\end{equation}
where the fermionic (antifermionic) fields $\psi\,(\psibar)$ are triplets in colour and flavour space, 
$M_0=m_0 1\!\!1$ is the bare mass matrix of $N_f=3$ degenerate quarks, and $D$ is the lattice Dirac operator for 
which we consider the $O(a)$-improved definition~\cite{Wilson:1975hf,Sheikholeslami:1985ij}
\begin{equation}
    D=D_{\rm w} + a D_{\rm sw}\,.
    \label{eq:Dirac}
\end{equation}
The operator $D_{\rm w}$ is the massless Wilson-Dirac operator
\begin{equation}
    D_{\rm w} = \frac{1}{2}\big\{\gamma_\mu(\nabstar\mu+\nab\mu)-a\nabstar\mu \nab\mu\big\}\,,
\end{equation}
where $\nabstar\mu,\nab\mu$ are the covariant lattice derivatives that act on the quark fields as follows:
\begin{align}
    a \nab\mu \psi(x) & =  U_\mu(x)\psi(x+ a \hat{\mu})-\psi(x)\; ,\nonumber \\[0.25cm]
    a \nabstar\mu \psi(x) & = \psi(x) - U_\mu(x- a \hat{\mu})^\dag\psi(x - a \hat{\mu})\,.
\end{align} 
The Sheikholeslami-Wohlert term is
\begin{equation}
    D_{\rm sw}\psi(x) = c_{\rm sw}(g_0) \frac{1}{4}
    \sigma_{\mu\nu} \widehat F_{\mu\nu}(x)\psi(x)
    \label{eq:DiracSW}
\end{equation}
where $\sigma_{\mu\nu}=\frac{i}{2}[\gamma_\mu,\gamma_\nu]$. The field $\widehat F_{\mu\nu}(x)$ is
the clover discretization of the field strength tensor given by
\begin{equation}
    \widehat  F_{\mu\nu}(x) = \frac{i}{8a^2}\big\{Q_{\mu\nu}(x)-Q_{\nu\mu}(x)\big\}\,
    \label{eq:CloverFmunu}
\end{equation}
with
\begin{multline}
	Q_{\mu\nu}(x) = U_\mu(x)U_\nu(x+a \hat{\mu})U_\mu(x+a \hat{\nu})^\dag U_\nu(x)^\dag\\
	+ U_\nu(x)U_\mu(x-a \hat{\mu}+a \hat{\nu})^\dag U_\nu(x-a \hat{\mu})^\dag U_\mu(x-a \hat{\mu})\\
	+ U_\mu(x-a \hat{\mu})^\dag U_\nu(x-a \hat{\mu}-a \hat{\nu})^\dag U_\mu(x-a \hat{\mu}-a \hat{\nu})U_\nu(x-a \hat{\nu})\\
	+ U_\nu(x-a \hat{\nu})^\dag U_\mu(x-a \hat{\nu})U_\nu(x+a \hat{\mu}-a \hat{\nu})U_\mu(x)^\dag\, .
\end{multline}
The coefficient $c_{\rm sw}(g_0)$ has been tuned non-perturbatively~\cite{Yamada:2004ja} so that the spectrum 
of the theory is free from $O(a)$-discretization effects~\cite{Sheikholeslami:1985ij,Luscher:1996sc}.

\section{Free-energy density in lattice perturbation theory}
\label{sec:Free-energy density in lattice perturbation theory}
In this Appendix we discuss the results of the free-energy density at one-loop order in lattice perturbation
theory, evaluated in the infinite spatial volume limit. These expressions, taken from  Ref.~\cite{DallaBrida:2020gux} (in particular Appendices E
and F), are used for the perturbative improvement of the entropy density computed non-perturbatively on the lattice.

\subsection{Preliminaries}
In this Section we consider lattice QCD with a generic number $N_c$ of colours, and $\Nf$ flavours of 
mass-degenerate $O(a)$-improved Wilson fermions (in the final numerical results we will then set $N_c=3$, $N_f=3$).
The lattice has compact size $L_0$ and spatial sizes $L_1$, $L_2$, $L_3$.

In the presence of shifted boundary conditions, the bosonic and the fermionic momenta in the first Brillouin zone are, respectively, given by
\begin{equation}
    p_0 = \frac{2\pi n_0}{L_0} - \sum_{k=1}^3 p_k\xi_k \,, \quad p_k = \frac{2\pi n_k}{L_k}
    \label{eq:BZ_boson}
\end{equation}
and
\begin{equation}
    p_0 = \frac{2\pi n_0}{L_0} + \frac{\pi}{L_0} - \sum_{k=1}^3 p_k\xi_k \,, \quad p_k = \frac{2\pi n_k}{L_k}\,,
\end{equation}
where $n_\mu = 0, 1, ..., L_\mu/a-1$.
In momentum space, the free gluonic propagator in the Feynman gauge reads
\begin{equation}
    D^{ab}_{\mu\nu}(p) = \frac{\delta_{ab}\delta_{\mu\nu}}{D_G(p)}\,, \quad 
    D_G(p) = \sum_{\mu=0}^3 \phat^2_\mu\,,
\end{equation}
with $\phat_\mu = \frac{2}{a}\sin\left(\frac{ap_\mu}{2}\right)$, while the free fermionic propagator is
\begin{equation}
S(p) = \frac{-i\gamma_\mu\pbar_\mu + m_0(p)}{D_F(p)}\,,\quad
  D_F(p) = \sum_{\mu=0}^3\pbar_\mu^2 + m_0^2(p)\,,
\end{equation}
where $\pbar_\mu = \frac{1}{a}\sin(ap_\mu)$, and
\begin{equation}
   m_0(p) = m_0 + \frac{a}{2}\sum_{\mu=0}^3 \phat_\mu^2\,.
\end{equation}
We use the notation $\int_{p_{_\vxi}}$ for the
integral
\footnote{Strictly speaking, the quantity in Eq.~\eqref{eq:mint} is a sum over the discrete momentum components. 
In the following we call these sums integrals because we are interested in the thermodynamic limit, where the sum 
becomes a sum of integrals over the spatial momenta.}
 over the momenta of a generic function $f(p)$
\begin{equation}
    \int_{p_{_\vxi}} f(p) = \frac{1}{L_0L_1L_2L_3}\sum_{n} f(p)\,,
    \label{eq:mint}
\end{equation}
without distinguishing between the bosonic and the fermionic cases since any potential ambiguity is resolved by the propagator that always appears in
the integrand. In the infinite spatial volume limit, the integral
in Eq.~\eqref{eq:mint} becomes
\begin{equation}
    \int_{p_{_\vxi}} f(p) \xrightarrow{L_k\to\infty} \frac{1}{L_0}\sum_{n_0} \int_{BZ} \frac{d^3\bs p}{(2\pi)^3}\, f(p)\,,
    \label{eq:integral_ivl_BZ}
\end{equation}
where $BZ$ stands for the Brillouin zone.

\subsubsection{Some relevant integrals}
We define here some integrals that appear in the lattice perturbative expansions,

\begin{align}
  B^{(0)} &= \int_{p_{_{\vxi}}} \frac{1}{D_G(p)}\,, \label{eq:B0} \\
  B^{(3)}_\mu &= \int_{p_{_\vxi}} \frac{\rmc_\mu(p)}{D_G(p)} \label{eq:B3mu}\,, \\
  F^{(4)}_{\mu\nu} &=\int_{p_{_{\vxi}}}  \frac{\bar p_\mu \cn(p)}{D_F(p)}\,, \label{eq:F4munu} \\
  F^{(5)}_{\mu} &=\int_{p_{_{\vxi}}}  \frac{m_0(p) \bar p_\mu}{D_F(p)}\,, \label{eq:F5mu} \\
  F^{(6)}_{\mu\nu}&=\int_{p_{_{\vxi}}}  \frac{m_0(p) \bar p_\mu \cn(p)}{D_F^2(p)}\,, \label{eq:F6munu} \\
  F^{(7)}_{\mu}&=\int_{p_{_{\vxi}}}  \frac{m_0^2(p) \bar p_\mu}{D_F^2(p)}\,, \label{eq:F7munu} \\
  F^{(8)} &= \int_{p_{_{\vxi}}} \frac{m_0(p)}{D_F(p)}\,, \label{eq:F8}
\end{align}
where $\rmc_\mu(p) = \cos(ap_\mu)$.

\subsubsection{Critical mass at one-loop order}
\label{sec:Critical_mass_at_one-loop_order}

In perturbation theory we define the critical mass $m_\rmcr$ with the same prescription
\footnote{This is alternative to the definition based on the quark self-energy, 
and the two differ by lattice artifacts.}
followed non-perturbatively, as the value of bare quark mass for which the PCAC mass vanishes 
in the Schr\"odinger functional setup~\cite{Kurth:2002rz,Panagopoulos:2001fn}. 
At one-loop order
\begin{multline}
    am_\rmcr = am_\rmcr^{(0)} + \delta am_\rmcr^{(0)} \\
             + g_0^2\left(am_\rmcr^{(1)} + \delta am_\rmcr^{(1,0)} + \Nf\delta am_\rmcr^{(1,1)}  \right)\,,
    \label{eq:amcr}
\end{multline}
where  $am_\rmcr^{(0)}=0$, and we define
\begin{align}
    am_\rmcr^{(1)} &= \frac{N_c^2-1}{\Nc}\, am_\rmcr^{(1,\Nc)}\,, \\
    \delta am_\rmcr^{(1,0)} &= \frac{N_c^2-1}{\Nc}\, \delta am_\rmcr^{(1,\Nc,0)}\,, \\
    \delta am_\rmcr^{(1,1)} &= \frac{N_c^2-1}{\Nc}\, \delta am_\rmcr^{(1,\Nc,1)}\,,
\end{align}
with~\cite{Panagopoulos:2001fn}
\begin{align}
   a m^{(1,N_c)}_\rmcr =& -0.16285705871085(1) \nonumber\\ 
                        &+c_{\rm sw}\, 0.04348303388205(10) \nonumber\\
                        &+ c_{\rm sw}^2 \, 0.01809576878142(1)\, .
   \label{eq:amc1Nc}
\end{align}
At this order, the improvement coefficient $c_{\rm sw}$ is either $0$ or $1$ for the unimproved or improved theory respectively.
The quantities $\delta am_\rmcr^{(0)}$, $\delta am_\rmcr^{(1,0)}$, $\delta am_\rmcr^{(1,1)}$ are the cutoff effects 
in the perturbative determination of the critical mass. 
The values for $\delta am_\rmcr^{(0)}$, $am_\rmcr^{(1)}+\delta am_\rmcr^{(1,0)}$ and $\delta am_\rmcr^{(1,1)}$ for 
$N_c=N_f=3$ and $c_{\rm sw}=1$ can be found in Table 7.1 of Ref~\cite{Kurth:2002rz}.
In our perturbative calculations at one-loop order, the proper chiral limit is thus obtained by setting in the integrals the bare quark mass
to the tree-level value of Eq.~\eqref{eq:amcr}.

\subsection{Free-energy density}
\label{ssec:Free-energy density}
We collect here the expression of the free-energy density in lattice perturbation theory at one-loop order. 
Results are taken from Appendix F of Ref.~\cite{DallaBrida:2020gux}.
At given $L_0/a$, we write the perturbative expansion as
\begin{equation}
    f_\vxi = f^{(0)} + g_0^2 f^{(1)}\,,
    \label{eq:fenergy_pt}
\end{equation}
where, for simplicity, we omit the dependence on $L_0/a$ and suppress the dependence of the perturbative coefficients on the shift  $\vxi$.
The tree-level coefficient reads
\begin{equation}
    f^{(0)} = (N_c^2-1)f^{G(0)} + \Nc\Nf f^{F(0)}
    \label{eq:fenergy_tree}
\end{equation}
with
\begin{equation}
f^{G(0)} = \int_{p_{_{\vxi}}}  \ln{\Big[a^2\, D_G(p)\Big]}\,,
\label{eq:fG0}
\end{equation}
\begin{equation}
f^{F(0)} = - 2\int_{p_{_{\vxi}}}  \ln{\Big[a^2 D_F(p)\Big]}\,.
\label{eq:fF0}
\end{equation}
At one-loop order we have
\begin{multline}
    f^{(1)} = (N_c^2-1)
    \Big\{ \Nc f^{G(1,\Nc)} + \frac{1}{\Nc}f^{G(1,\frac{1}{\Nc})} \\
    + \Nf \Big[f^{F(1,\Nf)} + \cF^{F1} + \cF^{F2} + \frac{\partial f^{F(0)}}{\partial am_0} \times \\
   \left(a\mcr^{(1,\Nc)} + \delta a\mcr^{(1,\Nc,0)} + \Nf\delta a\mcr^{(1,\Nc,1)}\right) \Big] \Big\}\,.
    \label{eq:fenergy_1loop}
\end{multline}
The gluonic coefficients are given by
\begin{multline}
    f^{G(1,N_c)} = \Big\{ (B^{(0)})^2  - \frac{1}{2} \sum_\sigma \Big[B^{(0)}-B_\sigma^{(3)} \Big]^2 \\
    +\! \frac{1}{2} a^2 K_1\! +
    \! \frac{1}{24} a^4 K_2\! -\! \frac{1}{2a^2} B^{(0)} \Big\}\,,
    \label{eq:fG1Nc}
\end{multline}
\begin{equation}
    f^{G(1,\frac{1}{\Nc})} = \frac{1}{2} \Big\{ \sum_\sigma \Big[B^{(0)}-B_\sigma^{(3)} \Big]^2 +\frac{1}{8a^4} \Big\}\,,
    \label{eq:fG11Nc}
\end{equation}
where we defined the integrals
\begin{equation}
    K_1 = \int_{p_{_{\vxi}};q_{_{\vxi}};k_{_{\vxi}}} \frac{\bar\delta(p+q+k)}{D_G(p)D_G(q)D_G(k)}
    \sum_\mu \hat p_\mu^2\, \hat q_\mu^2\, ,
    \label{eq:K1}
\end{equation}
\begin{equation}
    K_2 = \int_{p_{\vxi};q_{\vxi};k_{\vxi}} \frac{\bar\delta(p+q+k)}{D_G(p)D_G(q)D_G(k)}
    \sum_\mu  \hat p_\mu^2\, \hat q_\mu^2\, \hat k_\mu^2\,
    \label{eq:K2}
\end{equation}
with $\overline{\delta}(p) = a^4 \sum_x e^{ipx}$. The fermionic part is
\begin{multline}
    f^{F(1,N_f)} =
    B^{(0)} \left[ \frac{1}{a^2} - a \left(a m_0+ 4 \right) F^{(8)}\right]\!\! \\
    +\!\! \int_{q_{_{\vxi}};p_{_{\vxi}};k_{_{\vxi}}} \frac{\bar\delta(p-q-k)}{D_G(q)D_F(p)D_F(k)} \\
    \times \Big[a m_0(k)\sum_{\sigma} \bar r_\sigma \bar p_\sigma  + a m_0(p)\sum_{\sigma}\bar k_\sigma \bar r_\sigma \\
    -m_0(k) m_0(p)\sum_{\sigma}\rmc_{\sigma}(r)
    +\sum_{\sigma}\bar p_\sigma \bar k_\sigma \left(\rmc_{\sigma}(r)-3\right)\Big]\,,
    \label{eq:fF1Nf}
\end{multline}
where $r=p+k$. 
The terms $\cF^{F1}$ and $\cF^{F2}$ are the contributions from the $O(a)$-improvement in the Wilson action:
\begin{multline}
    {\cal F}^{F1}  =  - \frac{a c_{\rm sw}}{2}\int_{q_{_\vxi};p_{_{\vxi}};k_{_{\vxi}}} 
                      \frac{\bar\delta(p-q-k)}{D_G(q) D_F(p) D_F(k)} \\
      \times \Bigg\{a\! \sum_{\sigma \rho}\!\! \Big[(\bar p_\rho + \bar k_\rho) \bar q_\sigma
        (\bar p_\sigma \bar k_\rho - \bar k_\sigma \bar p_\rho)\Big] \\
     - \sum_\sigma \Big\{\bar q_\sigma \big[ m_0(k) \bar p_\sigma - m_0(p) \bar k_\sigma \big]
    \sum_{\rho\neq\sigma} \big[ \rmc_{\rho} (p) + \rmc_{\rho} (k) \big]\Big\}\Bigg\}\, ,
    \label{eq:cFF1}
\end{multline}
\begin{multline}
    {\cal F}^{F2} = \frac{a^2 c_{\rm sw}^2}{8}
    \int_{q_{_\vxi};p_{_{\vxi}};k_{_{\vxi}}}\!\!\!\!\!\!\,\,\,\
        \frac{\bar\delta(p-q-k)}{D_G(q) D_F(p) D_F(k)} \\
        \times \Bigg\{m_0(p) m_0(k) \sum_\sigma \Big\{{\bar q}^2_\sigma \Big[3 + \sum_{\rho\neq\sigma} \rmc_{\rho} (q) \Big]\!\Big\} \\ 
    + 2 \sum_{\sigma\rho} \bar k_\sigma  \bar q_\sigma \bar p_\rho \bar q_\rho 
    \Big(2\! -\! \rmc_\sigma(q)\!  +\! \sum_{\lambda\neq\rho}\rmc_\lambda (q) \Big) \\
    -\sum_{\sigma\rho} \bar k_\sigma  \bar p_\sigma {\bar q}_\rho^2 \Big(1\! -\! 2 \rmc_\sigma(q) +
    \!\sum_{\lambda\neq\rho}\rmc_\lambda (q) \Big) 
    \Bigg\}\,.
    \label{eq:cFF2}
\end{multline}
The last term in Eq.~\eqref{eq:fenergy_1loop} comes from the one-loop order correction to the critical mass $a\mcr$, 
with a coefficient given by the derivative in the bare mass of the tree-level fermionic free-energy, 
\begin{equation}
    \frac{\partial f^{F(0)}}{\partial m_0} = -4 F^{(8)}\,.
    \label{eq:dfdam0}
\end{equation}

\subsection{Infinite volume limit of gluonic contributions}
\label{sec:Infinite_volume_limit_of_gluonic_contributions}
To evaluate the lattice sums that appear in the computation of the perturbative coefficients in finite volume,
we have regularized the gluonic propagator by removing the zero-mode occurring when the momentum of the gluon vanishes.
The finite size effects of quantities containing gluonic loops are then power-like in the inverse spatial size of the box
--~we set $L_1=L_2=L_3=L$~-- with leading scaling of $O(1/L)$ or $O(1/L^3)$ as $L\to\infty$.
The $O(1/L)$ effects can be removed analytically, while the $O(1/L^3)$ ones can be extrapolated numerically with high accuracy
to the thermodynamic limit where any systematic effects introduced by the gluon zero-mode subtraction disappear.
The lattice sums required for the integrals at one-loop order have been performed in coordinate space~\cite{Luscher:1995zz} 
after using a efficient Fast Fourier Transform algorithm for the computation of the propagators.
We ran in quadruple precision because of large cancellations occurring in the lattice sums appearing in the various terms.

\subsubsection{Tree-level integrals}
\label{ssec:Analytic infinite volume limits}
The infinite spatial volume limit of the integrals $B^{(0)}$, $B^{(3)}_\mu$ and $f^{G(0)}$ is obtained by replacing the discrete sums over
$p_1, p_2, p_3$ with continuum integrations according to Eq.~\eqref{eq:integral_ivl_BZ}.
If the shift is chosen in the first spatial direction, the integration in $p_2$ and $p_3$ can be done analytically~\cite{Luscher:1995zz}
and we obtain the following expressions,
\begin{equation}
    B^{(0)}_\infty = \frac{1}{2 L_0}\sum_{n_0}
    \int_0^{\infty} \!\!\!\!dx \int_{BZ}\frac{dp_1}{2\pi}\, I_0(x)^2 \, e^{x\, C(p_0,p_1)}\,,
    \label{eq:B0ivl}
\end{equation}
\begin{equation}
    B^{(3)}_{\mu,\infty} = \frac{1}{2 L_0}\sum_{n_0}
    \int_0^{\infty} \!\!\!\!dx \int_{BZ}\frac{dp_1}{2\pi}\, I_0(x)^2 \, \rmc_\mu(p)\, e^{x\, C(p_0,p_1)}\,,
    \label{eq:B31ivl}
\end{equation}
for $\mu=0,1$, while for $k=2,3$ we have
\begin{equation}
    B^{(3)}_{k,\infty} = \frac{1}{2 L_0}\sum_{n_0}
    \int_0^{\infty} \!\!\!\!dx \int_{BZ}\frac{dp_1}{2\pi}\, I_0(x)I_1(x) \, e^{x\, C(p_0,p_1)}\,,
    \label{eq:B32ivl}
\end{equation}
and finally
\begin{multline}
  f^{G(0)}_\infty =  \frac{1}{a^2L_0}\sum_{n_0} \int_{BZ}\frac{dp_1}{2\pi}\, 
    \Bigg\{
    \ln\left[4-2C(p_0,p_1)\right] \\
    + \int_0^{\infty} \!\!\!\!dx\, \frac{\left[ e^{-2x}-I_0(x)^2 \right]}{x} \, e^{x\, C(p_0,p_1)} 
    \Bigg\}\,.
    \label{eq:fG0ivl}
\end{multline}
In these equations $C(p_0,p_1)=\rmc_0(p) + \rmc_1(p) - 4$ and $I_0$ and $I_1$ are the modified Bessel functions of the first
kind~\cite{abramowitz1965handbook}. 
The integrals over $p_1$ and $x$ can then be easily computed numerically with high accuracy.

\subsubsection{Integrals at one-loop order}
\label{ssec:Infinite volume extrapolations}
\begin{figure*}[t]
\begin{center}
\begin{minipage}{\columnwidth}
\includegraphics[width=\textwidth]{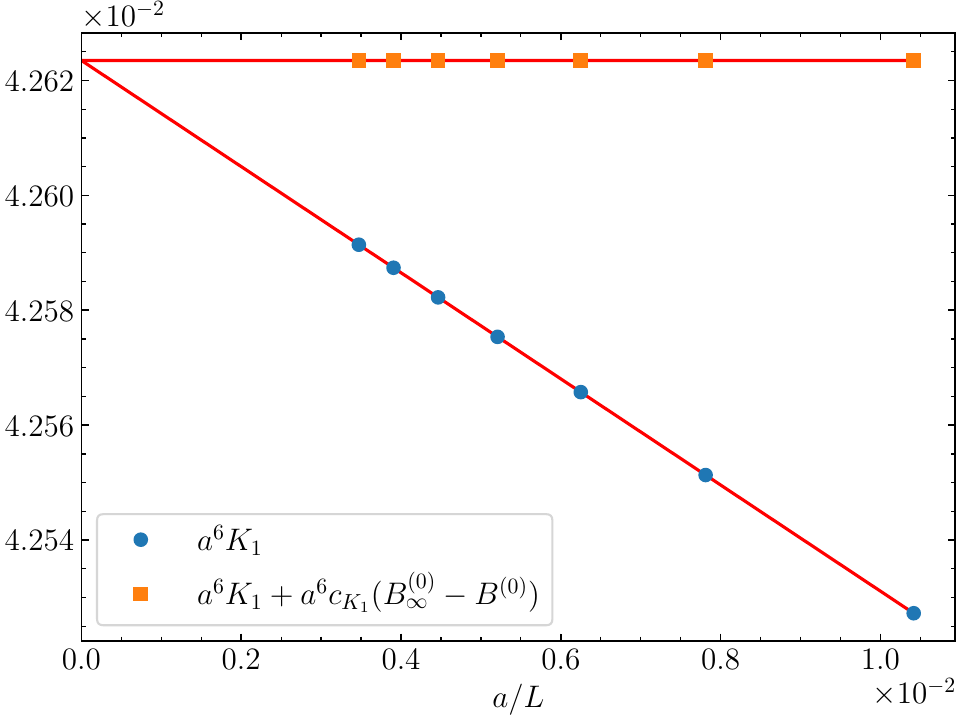}
\end{minipage}
\begin{minipage}{\columnwidth}
\includegraphics[width=\textwidth]{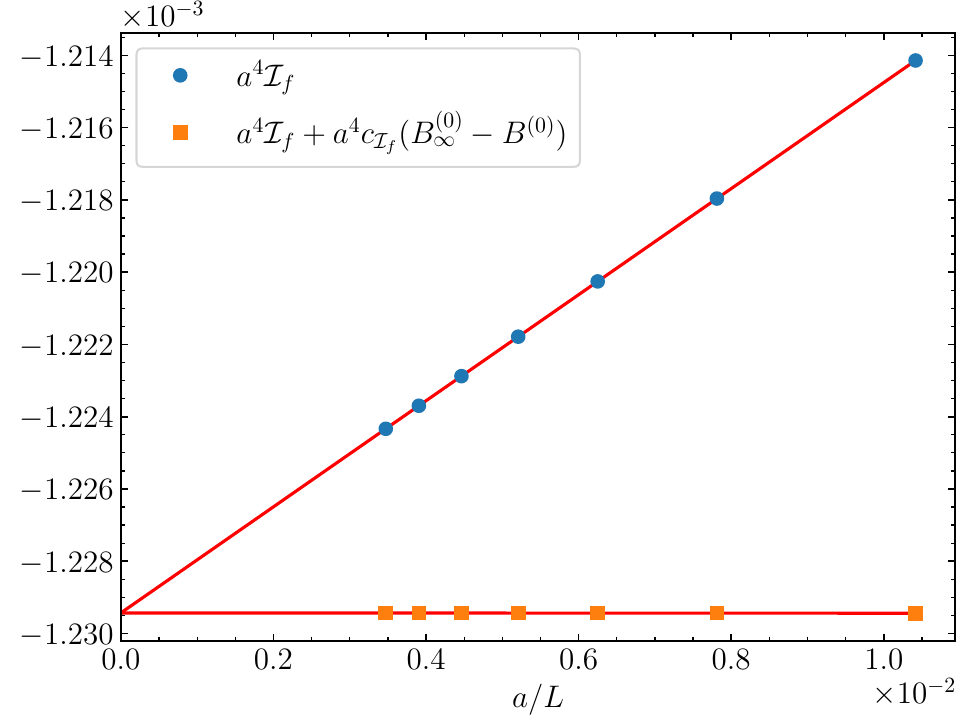}
\end{minipage}
\caption{Volume dependence of the integrals $K_1$ and ${\cal I}_f$ respectively defined in Eqs.~\eqref{eq:K1} and~\eqref{eq:If},
for the representative case $L_0/a=6$ and shift $L_0\vxi=(8a,0,0)$.
The spatial size $L/a$ of the lattice ranges between $96$ and $288$.
For both integrals the comparison with the subtracted definition is shown, see main text for the details.
Red lines represent the extrapolations to the thermodynamic limit.}
\label{fig:K1Ifivl}
\end{center}
\end{figure*}

We discuss here the computation of the integrals $K_1$, $K_2$ in the thermodynamic limit. These integrals, defined in 
Eqs.~\eqref{eq:K1} and~\eqref{eq:K2}, enter in the one-loop order gluonic contribution $f^{G(1,\Nc)}$ of Eq.~\eqref{eq:fG1Nc}.
The integral $K_1$ has leading finite volume effects of order $O(1/L)$. In the small momentum limit of $k$, the $\bar\delta$ function leads
to $q\sim -p$ and we have the effective behaviour
\begin{equation}
  K_1 \sim \int_{k_{\vxi}}\frac{1}{D_G(k)}\int_{p_\vxi}\frac{\sum_\mu \hat{p}^4_\mu}{D_G^2(p)} = B^{(0)}\, c_{K_1}
\end{equation}
whose leading $O(1/L)$ finite size effects are given by $B^{(0)}$. The second integral in r.h.s., denoted by $c_{K_1}$, is
the residual at the pole $k = 0$ of the integrand function of $K_1$. It is now convenient to consider the subtracted integral
\begin{equation}
    K_1'= K_1 + c_{K_1}\left(B^{(0)}_\infty - B^{(0)} \right)\,,
    \label{eq:K1sub}
\end{equation}
which attains the same value of $K_1$ in the thermodynamic limit but with the faster convergence rate $O(1/L^3)$, as
the leading $O(1/L)$ finite volume corrections of $K_1$ have been removed. 
Moreover, the integral appearing in the coefficient $c_{K_1}$ can be easily computed with high accuracy. 
  
In left panel of Figure~\ref{fig:K1Ifivl} we compare the approach to the thermodynamic limit of $K_1'$ 
(orange square markers) and $K_1$ (blue circle markers). 
The system size in the spatial directions increases from $L/a=96$ up to $L/a=288$.
For all practical purposes we can quote as the infinite volume limit the value of $K_1'$ at the largest volume, 
whose discrepancy with respect to the extrapolation is well below the permille level.

The leading finite size corrections of the integral $K_2$ are of order $O(1/L^3)$ and they are negligible 
at the largest size $L/a=288$ that we have considered. Thus, we take that value as the thermodynamic limit.

\subsection{Infinite volume limit of fermionic contributions}
\label{ssec:Infinite volume limit of fermionic contributions}
We have extrapolated to the thermodynamic limit the tree-level fermionic free-energy density $f^{F(0)}$, 
Eq.~\eqref{eq:fF0}, and the one-loop order mass counterterm, Eq.~\eqref{eq:dfdam0}, 
by increasing $L/a$ from $96$ to $288$ at fixed $L_0/a$.
The convergence to the infinite volume limit is very fast because, at asymptotically high temperatures, fermions 
develop a thermal mass $M_{\rm th}\sim \pi T$ and finite volume effects are exponentially suppressed as $e^{-M_{\rm th}L}$~\cite{Giusti:2012yj}.
At the largest volume, finite size effects are well below the permille level, and in practice the results can be
considered as the values in the thermodynamic limit. 

The fermionic contributions at one-loop order have power-like finite size effects in $1/L$ because of integrable singularities related to
virtual gluons in the loops. For example, in the fermionic coefficient $f^{F(1,\Nf)}$ of Eq.~\eqref{eq:fF1Nf}, we have the integral
\begin{multline}
    {\cal I}_f \equiv
    \int_{q_{_\vxi};p_{_\vxi};k_{_\vxi}}\frac{\bar\delta(p-q-k)}{D_G(q)D_F(p)D_F(k)} \Big[a m_0(k)\sum_{\sigma} \bar r_\sigma \bar p_\sigma \\
    + a m_0(p)\sum_{\sigma}\bar k_\sigma \bar r_\sigma -m_0(k) m_0(p)\sum_{\sigma}\rmc_{\sigma}(r)\\     
    +\sum_{\sigma}\bar p_\sigma \bar k_\sigma \left(\rmc_{\sigma}(r)-3\right)\Big]\,,\qquad \qquad \qquad \quad
    \label{eq:If}
\end{multline}
whose integrand has a pole when the gluonic momentum vanishes $q\to0$ and $p\sim k$. Following the same strategy as for the computation
of $K_1$ in the previous Section, we have introduced the subtracted integral
\begin{equation}
    {\cal I}_f'= {\cal I}_f + c_{{\cal I}_f}\left(B^{(0)}_\infty - B^{(0)}\right)\,,
    \label{eq:Ifsub}
\end{equation}
where $c_{{\cal I}_f}$ is the residual at the pole $q=0$ of the integrand function of ${\cal I}_f$, 
\begin{multline}
    c_{{\cal I}_f} = \int_{p_\vxi}\frac{1}{D_F^2(p)}
    \Big[ 2am_0(p)\sum_\sigma\overline{(2p)}_\sigma \overline{p}_\sigma \\
    - m_0^2(p)\sum_\sigma\rmc_\sigma(2p) + \sum_\sigma\overline{p}^2_\sigma\left(\rmc_\sigma(2p)-3\right)\Big]\,,
    \label{eq:cIf}
\end{multline}
and can be easily computed numerically with high accuracy.
The integrals ${\cal I}_f'$ and ${\cal I}_f$ have the same thermodynamic limit, 
but the leading finite volume effects of the former are of $O(1/L^3)$ as the leading ones of the 
latter, of $O(1/L)$, have been subtracted. 
The data for ${\cal I}_f'$ are represented in the right panel of Figure~\ref{fig:K1Ifivl} 
(orange square markers) in comparison with the results for ${\cal I}_f$ (blue circle markers).
As before, the residual finite size effects after the subtraction are negligible and we can take
the value at the largest size $L/a=288$ as the thermodynamic limit.

The same holds for the infinite volume extrapolation of the integrals $\cF^{F1}$, $\cF^{F2}$ defined respectively 
in Eqs.~\eqref{eq:cFF1} and~\eqref{eq:cFF2}, whose finite size corrections are already of $O(1/L^3)$. 

\section{Choice of the shift parameter}
\label{app:Choice of the shift parameter}
In this Appendix we discuss the choice of the shift parameter $\vxi$ that defines the boundary conditions 
for the gauge and fermionic fields along the compact direction, see Eqs.~\eqref{eq:shBCs_A} and~\eqref{eq:shBCs_psi}.
To this aim we consider the following generalization of Eq.~\eqref{eq:entropy_lattice},
\begin{equation}
    \frac{s}{T^3} = \frac{(1+\vxi^2)^3}{\vxi\cdot\bs{v}} \, L_0^4 \,
                    \frac{L_0}{2a} \left(f_{\vxi+\frac{a}{L_0}\bsv} - f_{\vxi-\frac{a}{L_0}\bsv}\right)\,, 
    \label{eq:s_DfDxi_v}
\end{equation}
where the spatial vector $\bsv$ defines the direction and the magnitude of the finite difference. 
In the continuum limit, this finite difference becomes the directional derivative of $f_\vxi$ along $\bsv$.
Different values of $\vxi$ and $\bsv$ lead to different discretizations of the entropy density, with different cutoff effects
and relative error: a convenient choice should minimize both of them.
We have considered the following set of shift vectors,
\begin{multline}
    \vxi=\big\{(1/2, 0, 0)\,,\, (1,0,0)\,, \\
         (1/2, 1/2, 0)\,, \, (1, 1, 0)\,, \, (2, 0, 0)\big\}\,,
    \label{eq:xi_candidates}
\end{multline}
so as to explore the trend of discretization effects and relative error of $s/T^3$ for increasing norm of $\vxi$ and for 
shifts in one or two directions.
For shifts in the first direction only, we have taken $\bsv=(1,0,0)$ (one-point finite difference in Eq.~\eqref{eq:s_DfDxi_v}) or
$\bsv=(2,0,0)$ (two-point finite difference). 
For non-zero shifts in the first and second direction, we have considered $\bsv=(1,1,0)$ (one-point simultaneous finite difference in 
both directions).

\subsection{Discretization effects}
\label{ssec:Discretization effects}
\begin{figure*}[t]
\begin{center}
\begin{minipage}{\columnwidth}
\includegraphics[width=\textwidth]{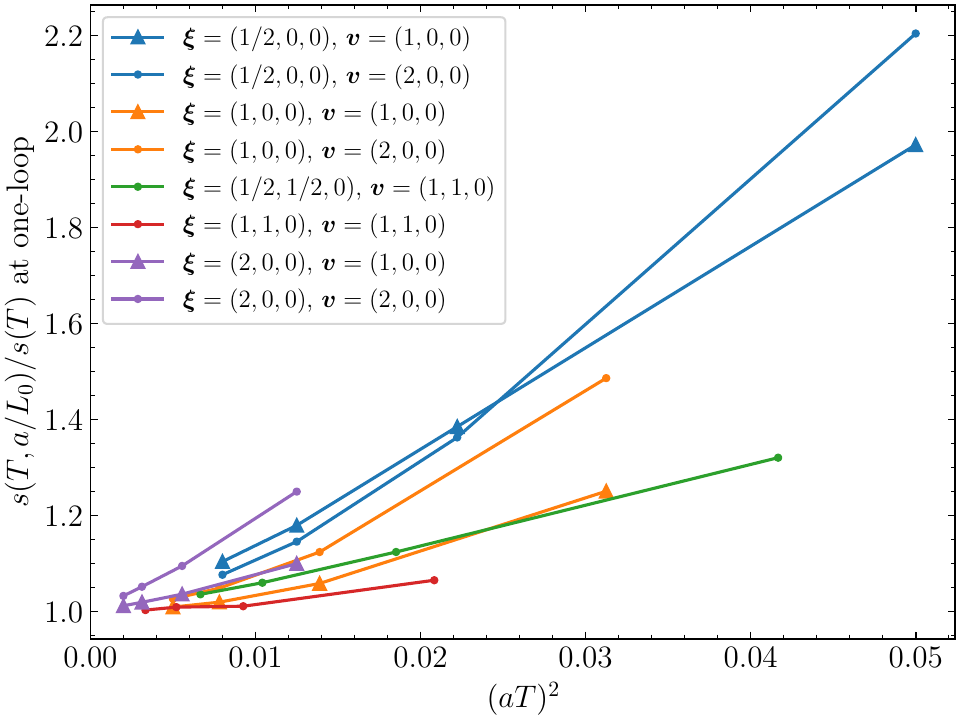}
\end{minipage}
\begin{minipage}{\columnwidth}
\includegraphics[width=\textwidth]{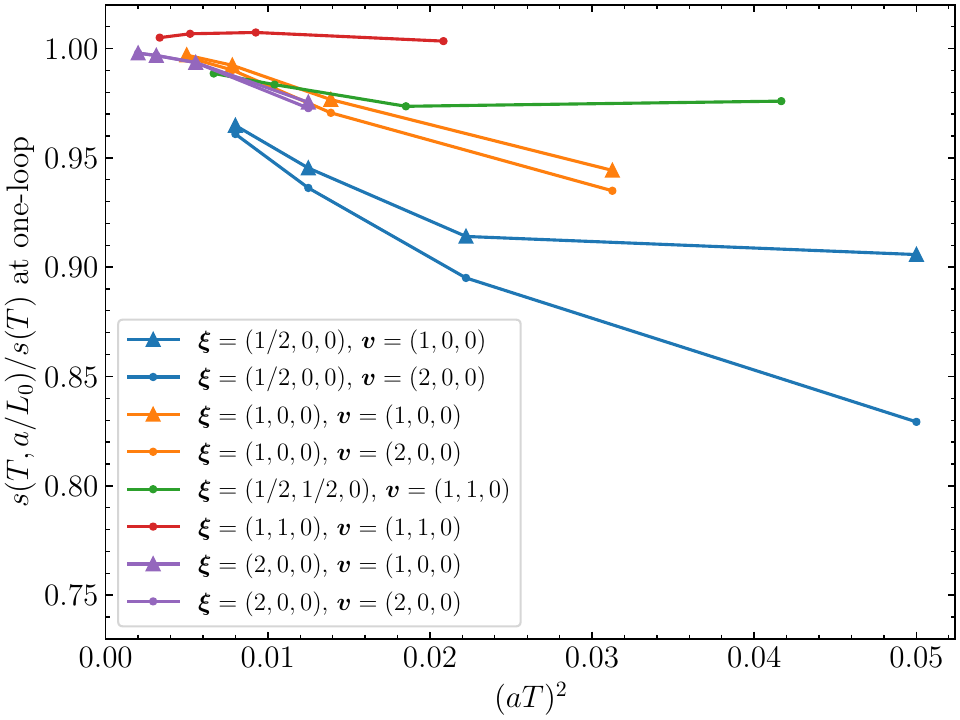}
\end{minipage}
\caption{Left: results in lattice perturbation theory at one-loop order of $s/T^3$, normalized to the continuum limit, 
         for some choices of $\vxi$ and $\bsv$.
         In the perturbative expressions we have used the SF coupling $\overline g_{\rm SF}^2(1/L_0)$ at the temperature $T_1$.
         The lines connecting the points are drawn to guide the eye.
         Right: same as left panel, but with tree-level improved data according to Eq.~\eqref{eq:pt_improvement}.}
\label{fig:ceff_s_lpt}
\end{center}
\end{figure*}
\begin{figure*}[t]
\begin{center}
   \begin{minipage}{\columnwidth}
   \includegraphics[width=\textwidth]{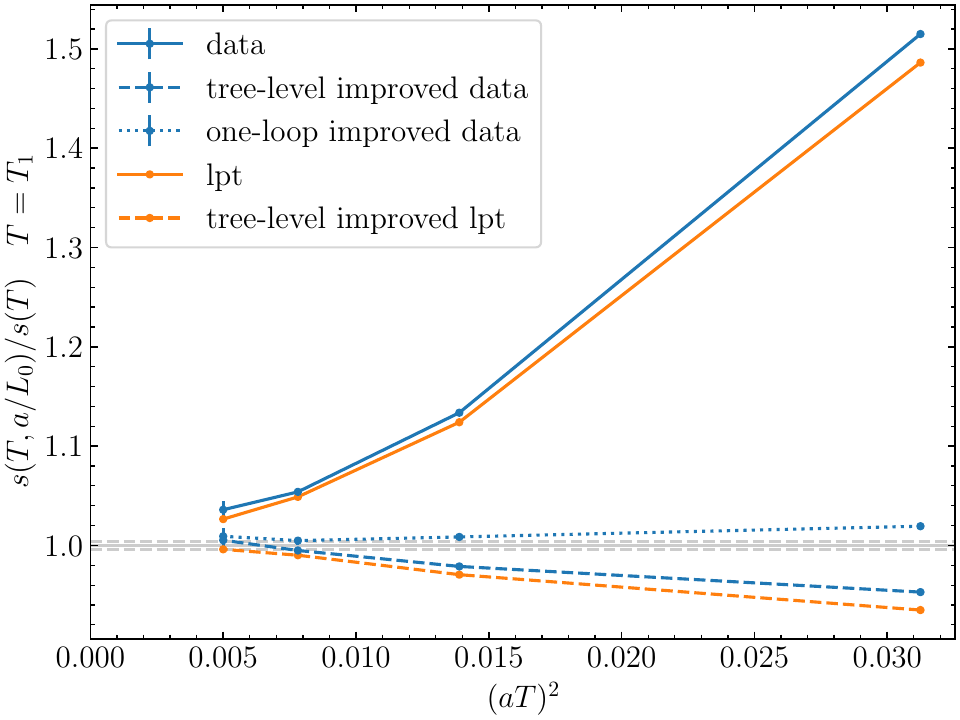}
   \end{minipage}
   \begin{minipage}{\columnwidth}
   \includegraphics[width=\textwidth]{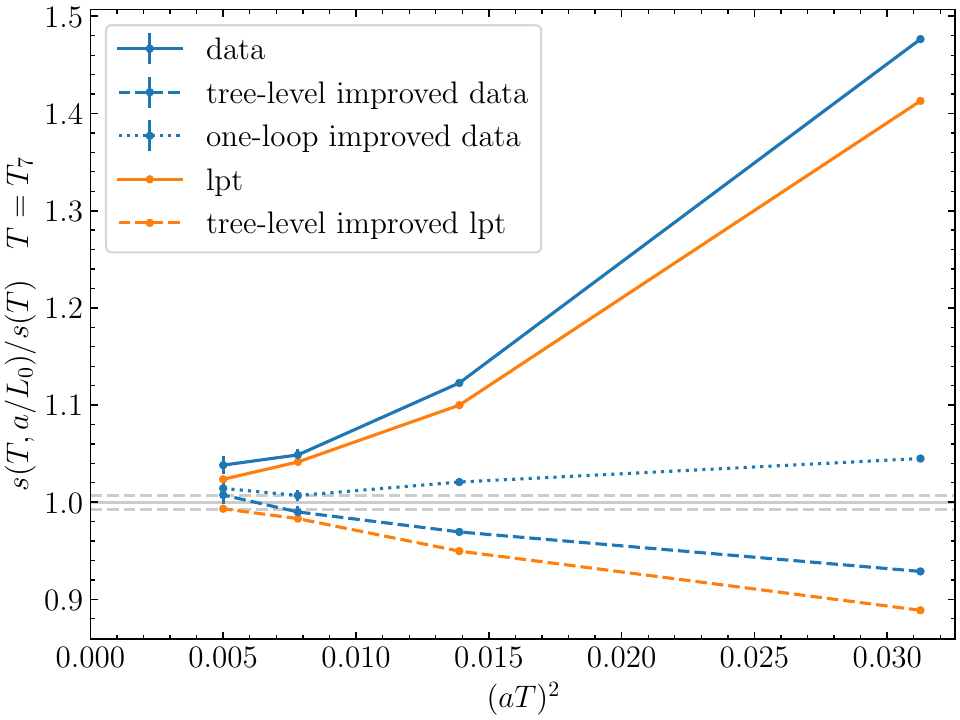}
   \end{minipage}
   \caption{Non-perturbative results of $s/T^3$ for $\vxi=(1,0,0)$ and $\bsv=(2,0,0)$,
            in comparison with lattice perturbation theory (lpt) at one-loop order.
            Non-perturbative and perturbative results are normalized to their respective continuum limits.
            The left panel shows data at the temperature $T_1$, the right panel at the temperature $T_7$.
            The grey dashed lines represent the uncertainty of the extrapolated non-perturbative continuum limit.
            The lines connecting the points are drawn to guide the eye.
            }
   \label{fig:ceff_s}
\end{center}
\end{figure*}
We have studied how discretization effects depend on the values of $\vxi$ and $\bsv$ using lattice perturbation theory at one-loop order.
Figure~\ref{fig:ceff_s_lpt} shows the entropy, computed as in Eq.~\eqref{eq:s_DfDxi_v}, normalized to the continuum result at one-loop order, 
for several values of the parameters under investigation.
Data with $\vxi=(1/2,0,0)$ have the largest cutoff effects, data with $\vxi=(1,1,0)$ have the smallest.
Most of the discretization effects can be actually removed with a tree-level improvement as in Eq.~\eqref{eq:pt_improvement} 
(with $g=0$), and the remaining lattice artifacts are of the order of a few percent 
for all the considered values of the shift but for $\vxi=(1/2,0,0)$, which still shows a $10\%-15\%$ deviation from the continuum limit
at the coarsest point.
Furthermore, at a given $\vxi$, the tree-level improvement removes almost all the discrepancy between the one-point and two-point
discretizations of the finite difference.
The validity of these results at the non-perturbative level is confirmed by Figure~\ref{fig:ceff_s}, where
the cutoff effects of the non-perturbative data for $\vxi=(1,0,0)$, $\bsv=(2,0,0)$ compare very well with the perturbative ones at the temperatures $T_1$ and $T_7$. Similar conclusions hold at all other temperatures.

This study shows that the discretization effects of $s/T^3$ can be removed very efficiently through the perturbative 
improvement, after which all the considered values of $\vxi$ and $\bsv$, with the only exception of $\vxi=(1/2,0,0)$, 
are comparable in terms of residual lattice artifacts.

\subsection{Relative error}
In the presence of shifted boundary conditions, the temperature of the thermal system depends on the shift through the relation 
$T^{-1}=L_0\sqrt{1+\vxi^2}$.
At a given $L_0$, the temperature changes by less than a factor 2 with respect to the reference case $\vxi=(1,0,0)$ when
the shift is varied within the values listed in Eq.~\eqref{eq:xi_candidates}.
Since the entropy density increases slowly from 3 GeV to 165 GeV (about $3\%$, see Table~\ref{tab:fits_comparison}), 
at these temperatures and at a given $L_0$ we can assume $s/T^3$ to be 
approximately constant when the shift is changed 
among the values under investigation.
Using this assumption in Eq.~\eqref{eq:s_DfDxi_v}, we get the scaling
\begin{equation}
    L_0^4\left(f_{\vxi+\frac{a}{L_0}\bsv} - f_{\vxi-\frac{a}{L_0}\bsv}\right) \propto \frac{\vxi\cdot\bs{v}}{(1+\vxi^2)^3}
\end{equation}
for the finite difference of the free-energy, while the related variance is expected to be mostly independent on the shift. This leads to the relation
\begin{equation}
    \frac{\sigma[s]}{s} \propto \frac{(1+\vxi^2)^3}{\vxi\cdot\bsv}
    \label{eq:erel_s_DfDxi_v}
\end{equation}
for the relative accuracy of the entropy density.
A direct non-perturbative check of this scaling would be computationally very demanding.
Similarly to Subsection~\ref{ssec:Finite volume effects}, we have instead verified the analogous relation for the energy-momentum tensor,
\begin{equation}
   \frac{\sigma\big[\corr{T^{R,\{6\}}_{0k}}_\vxi\big]}{\corr{T^{R,\{6\}}_{0k}}_\vxi} \propto \frac{(1+\vxi^2)^3}{\xi_k}\,,
   \label{eq:erel_T0k}
\end{equation}
coming from Eq.~\eqref{eq:s_T0k6}.
We have simulated $8\times48^3$ lattices at the temperature $T_1$ and at the values of shift given in Eq.~\eqref{eq:xi_candidates}. 
For each ensemble, the measured expectation values of the components $T^{G,\{6\}}_{01}$ and $T^{F,\{6\}}_{01}$, defined in 
Eq.~\eqref{eq:T6R}, are reported in Table~\ref{tab:s_T0k}.
The data confirm the assumption that the variance is mostly independent on the shift.
Furthermore, the relative errors are well described by the expected scaling Eq.~\eqref{eq:erel_T0k}.
\begin{table}[t]
    \centering
    \begin{tabular}{|c|cc|c|}
    \hline
    $\vxi$ & $L_0^4\,\corr{T^{G,\{6\}}_{01}}_\vxi$ & $L_0^4\,\corr{T^{F,\{6\}}_{01}}_\vxi$ & Eq.~\eqref{eq:erel_T0k} \\
    \hline
	$(1/2,   0,   0)$ & $   -1.60(24)$ & $    -3.50(8)$ & $3.91$ \\
	$(  1,   0,   0)$ & $   -0.87(22)$ & $    -1.60(9)$ & $8.00$ \\
	$(1/2, 1/2,   0)$ & $   -0.92(19)$ & $    -1.88(9)$ & $6.75$ \\
	$(  1,   1,   0)$ & $   -0.18(21)$ & $   -0.40(10)$ & $27.00$ \\
	$(  2,   0,   0)$ & $   -0.17(28)$ & $    -0.21(9)$ & $62.50$ \\
    \hline
 \end{tabular}
    \caption{Results for the energy-momentum tensor from $8\times 48^3$ lattices at the temperature $T_1$. 
             Each ensemble has 100 measurements.
             The relative error of the matrix elements is expected to scale as the numbers in the last column.}
    \label{tab:s_T0k}
\end{table}
\begin{table}[t]
    \centering
    \begin{tabular}{|c|c|c|}
        \hline
        $\vxi$ & $\bsv$ & Eq.~\eqref{eq:erel_s_DfDxi_v} \\
        \hline
		$(1/2,   0,   0)$ & $(1, 0, 0)$ & $3.91$ \\
		$(1/2,   0,   0)$ & $(2, 0, 0)$ & $1.95$ \\
		$(  1,   0,   0)$ & $(1, 0, 0)$ & $8.00$ \\
		$(  1,   0,   0)$ & $(2, 0, 0)$ & $4.00$ \\
		$(1/2, 1/2,   0)$ & $(1, 1, 0)$ & $3.38$ \\
		$(  1,   1,   0)$ & $(1, 1, 0)$ & $13.50$ \\
		$(  2,   0,   0)$ & $(1, 0, 0)$ & $62.50$ \\
		$(  2,   0,   0)$ & $(2, 0, 0)$ & $31.25$ \\
        \hline
    \end{tabular}
    \caption{Expected scaling of $\sigma[s]/s$.}
    \label{tab:s_DfDxi_v}
\end{table}

Given that the assumptions underlying Eq.~\eqref{eq:erel_s_DfDxi_v} are well supported by the non-perturbative results, 
we have collected in Table~\ref{tab:s_DfDxi_v} the expected scaling for the relative error of $s/T^3$ at the different choices
of $\vxi$ and $\bsv$ under investigation.
The two-point finite difference leads to halved relative errors with respect to the one-point one.
Since any additional discretization effects can be removed efficiently with the tree-level improvement 
(see Subsection~\ref{ssec:Discretization effects}), the choice $\bsv=(2,0,0)$ is preferable to $\bsv=(1,0,0)$.
On the other hand, the relative error increases with the norm of the shift, at the point that 
the shifts $\vxi=(1,1,0)$ and $\vxi=(2,0,0)$ are severely penalized.
We may also exclude $\vxi=(1/2, 0, 0)$, which leads to larger cutoff effects compared to the other shifts.
Choices like $\vxi=(1/2,1/2,0)$ with $\bsv=(1,0,0)$ would lead to twice the relative error of $\bsv=(1,1,0)$, 
and if we choose instead $\bsv=(2,0,0)$ the signal is the same as $\bsv=(1,1,0)$ while the cutoff effects are expected to be worse.
The relevant candidates are thus $\vxi=(1,0,0)$ with $\bsv=(2,0,0)$ or $\vxi=(1/2,1/2,0)$ with $\bsv=(1,1,0)$.
After the tree-level improvement, their residual cutoff effects are comparable, while the relative error on the entropy with 
$\vxi=(1/2,1/2,0)$ is about $15\%$ less.
At present, our simulation code allows only for even values of $L_0\vxi$ in every spatial direction.
We have therefore chosen $\vxi=(1,0,0)$ for which all the resolutions $L_0/a=4,6,8,10$ can be simulated, 
also considering that lattices with $L_0/a>10$ are computationally very demanding.

\section{Systematics from the numerical quadratures}
\label{sec:Systematics from the numerical quadratures}
In this Appendix we discuss the choice of the integration scheme for the integral in the bare subtracted mass 
defined in Section~\ref{sec:Numerical computation}. 
We have used lattice perturbation theory at tree-level as guidance, since at this perturbative order we can calculate 
both sides of the relation
\begin{equation}
   \frac{\Delta (f_\vxi - f_\vxi^\infty)}{\Delta\xi_k}
   = -\int_0^\infty dm_q \frac{\Delta\corrno{\psibar\psi}_\vxi^{m_q}}{\Delta\xi_k}\,,
   \label{eq:sys_pt}
\end{equation}
and thus assess the systematic effects induced by the computation of the integral by means of numerical quadratures.

\begin{figure*}[t]
\begin{center}
\begin{minipage}{\columnwidth}
\includegraphics[width=\textwidth]{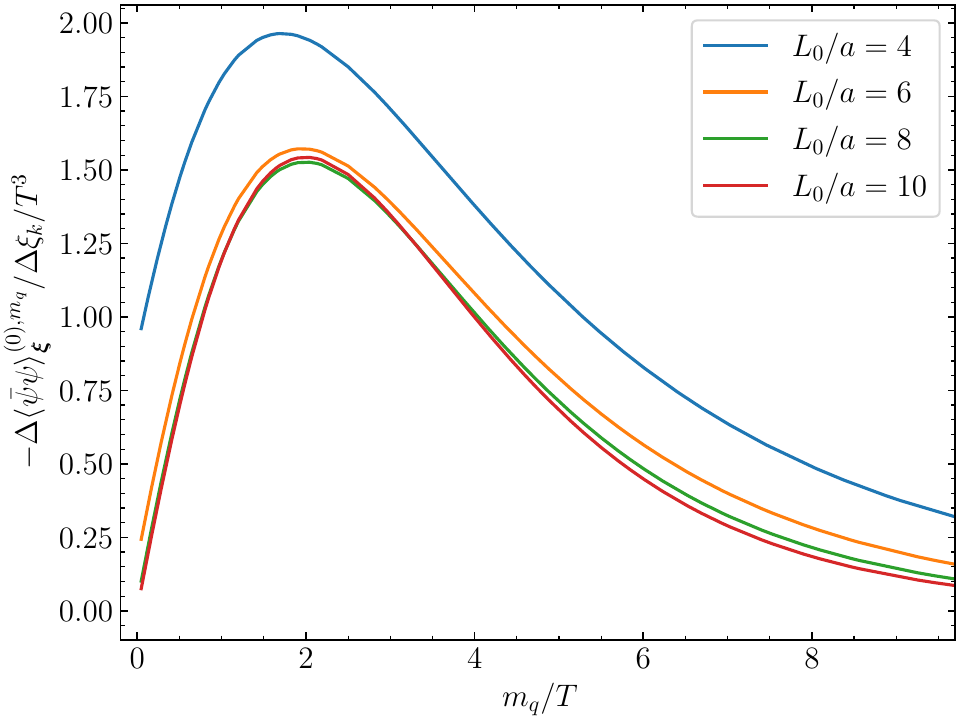}
\end{minipage}
\begin{minipage}{\columnwidth}
\includegraphics[width=\textwidth]{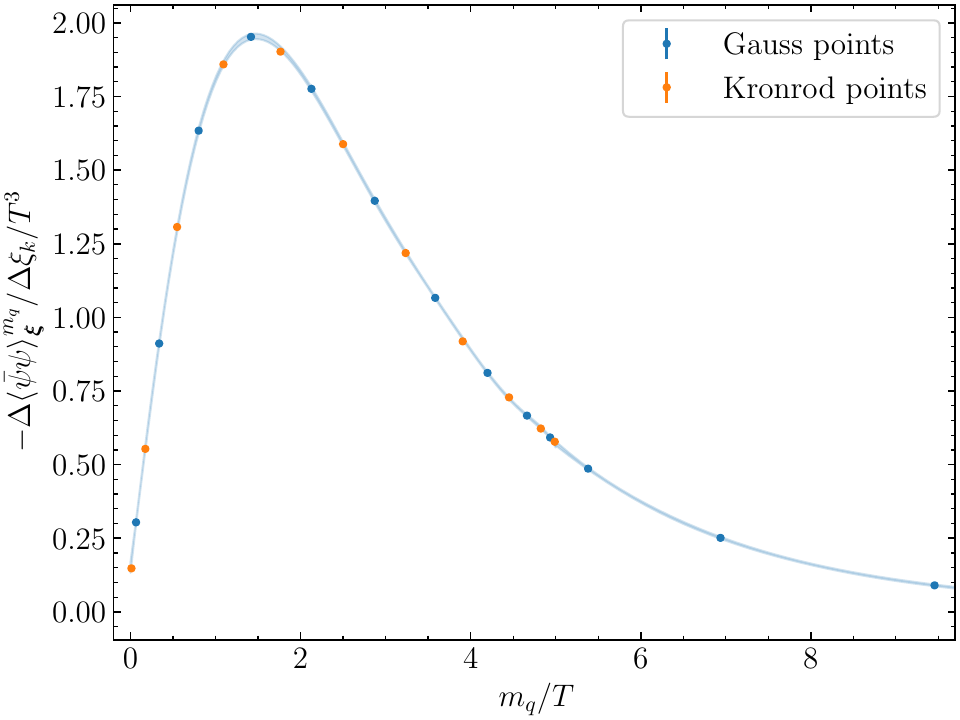}
\end{minipage}
\caption{Left: discrete derivative of the expectation value of the scalar density with respect to the shift, computed at 
tree-level in lattice perturbation theory at several values of $m_q/T$. 
Right: non-perturbative results at inverse bare coupling $6/g_0^2=8.2170$ computed on $6\times144^3$ lattices.}
\label{fig:systematics_Gauss_quadrature}
\end{center}
\end{figure*}

The left panel of Figure~\ref{fig:systematics_Gauss_quadrature} shows that the integrand function, at tree-level in lattice 
perturbation theory, has a very smooth dependence on the subtracted quark mass $m_q/T$.
The mild peaked shape suggests to split the numerical integration in three intervals for increasing bare mass, such that
the first segment is about $80\%$ of the total integral, the second segment about $20\%$ and the last part is
at most about equal to the target absolute error of the full integral.
A numerical exploration of the problem led to the integration scheme reported in Eq.~\eqref{eq:integral_mq_split}.
When the latter is applied to Eq.~\eqref{eq:sys_pt} at tree-level, the resulting systematic effects from the
numerical quadratures are at most $1/10$ of the target statistical accuracy.
A side advantage of splitting the integration in several domains is that, if needed, the numerical accuracy of each interval 
can be improved independently from the others by using, for instance, higher order quadratures.

The comparison of the right panel of Figure~\ref{fig:systematics_Gauss_quadrature} with the left panel
reveals that the non-perturbative behaviour of the integrand function is very similar to its perturbative counterpart.
The chosen integration scheme is thus expected to be appropriate also for the non-perturbative case.

We have explicitly checked on the non-perturbative results that the systematic effects from the numerical quadratures are 
negligible within the statistical error by improving the computation of the integral in the domain $\{0\leq m_q/T\leq 5\}$ at 
the bare parameters $6/g_0^2=8.2170$, $L_0/a=6$ (corresponding to the temperature $T_2$) using the Gauss-Kronrod quadrature rule.
Any systematic effects from the other integration domains are expected to be completely subdominant because the integrand function 
for $m_q/T\geq 5$ is monotonically decreasing, and because the bulk of the integral comes from the first domain.
The $n$-point Gauss quadrature integrates exactly polynomials up to order $2n-1$. The Gauss-Kronrod quadrature adds $n+1$ points 
to the Gauss quadrature, and integrates exactly polynomials up to order $3n+1$~\cite{Rabinowitz1980TheED}.
The points of the Gauss quadrature are shown in the right panel of Figure~\ref{fig:systematics_Gauss_quadrature} as blue markers,
and the blue shadowed band is a polynomial interpolation of them.
We have generated on the lattice the additional $11$ points (orange markers) for the $n=10+11$ Gauss-Kronrod rule. 
The error bars are smaller than the markers, and the orange points are compatible within about $0.5\sigma$ with the blue 
shadowed interpolation. 
\begin{table}[h]
    \centering
    \begin{tabular}{|c|c|}
        \hline
        Quadrature & $s/T^3$ \\
        \hline
         Gauss &  $22.77(5)$\\
         Gauss-Kronrod &  $22.76(5)$\\
         \hline
    \end{tabular}
    \caption{Results for the normalized entropy density at inverse bare coupling squared $6/g_0^2=8.2170$ computed on $6\times144^3$ lattices,
    using either the Gauss quadrature or the Gauss-Kronrod quadrature. See main text for the details.}
    \label{tab:Gauss-Kronrod}
\end{table}

The comparison between the two numerical integrations is in Table~\ref{tab:Gauss-Kronrod}, where we give the full result 
for $s/T^3$ when
the integral in the bare quark mass is computed as in Eq.~\eqref{eq:integral_mq_split} using either the $n=10$ Gauss rule or 
the $n=10+11$ Gauss-Kronrod rule for the domain $\{0\leq m_q/T\leq 5\}$. The statistical accuracy of the points have been tuned 
so that the two results have comparable absolute errors. 
The tiny difference is several times smaller than one combined standard deviation.
Furthermore, the shape of the non-perturbative integrand function depends very mildly on the temperature and on the value of 
$L_0/a$. The systematic effects from the Gauss quadratures are thus negligible for all the temperatures and lattice spacings
that we have simulated.

\section{Details on the lattice QCD simulations}
\label{app:Details on lattice QCD simulations}
We have simulated lattice QCD with $N_f=3$ degenerate flavours of quarks by using the HMC algorithm.
Our code is based on the package \texttt{openQCD-1.6}~\cite{openQCD,Luscher:2012av}, modified for including 
shifted boundary conditions~\cite{DallaBrida:2017sxr,DallaBrida:2021ddx}.
The doublet of up and down quarks have been simulated with an optimized twisted-mass Hasenbusch
preconditioning of the quark determinant~\cite{Hasenbusch:2001ne}.
The strange quark has been included with the RHMC algorithm~\cite{Kennedy:1998cu,Clark:2003na}.
We employed even-odd preconditioning for all the three flavours~\cite{Luscher:2010ae}.
The solution of the Dirac equation in the molecular dynamics trajectory was 
performed with a standard conjugate gradient with chronological inversion.
\begin{figure*}[t]
\begin{center}
\begin{minipage}{\columnwidth}
\includegraphics[width=\textwidth]{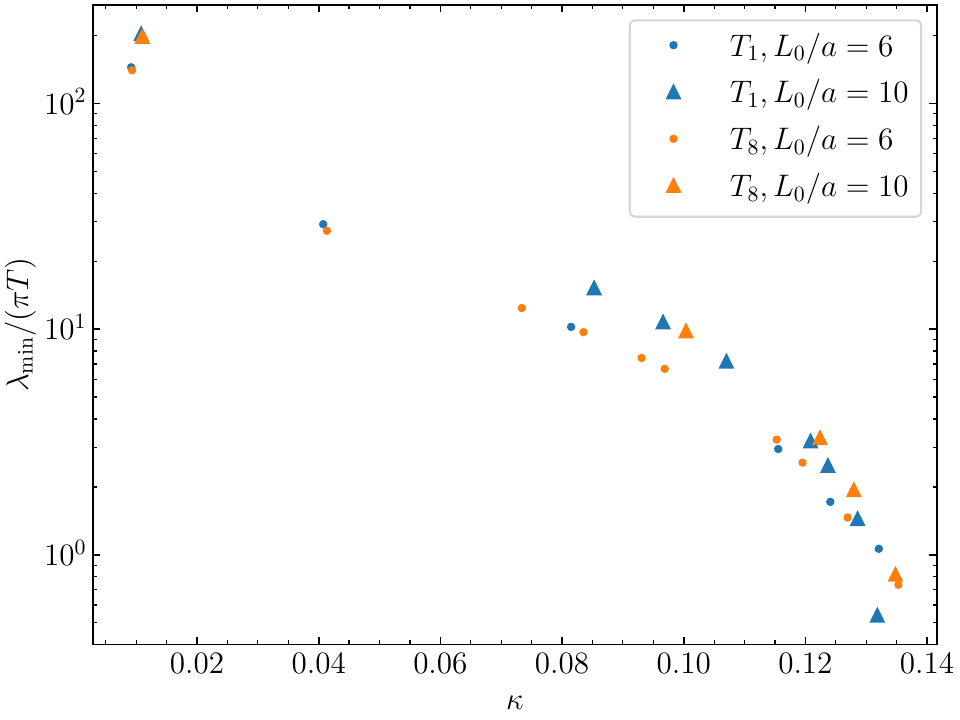}
\end{minipage}
\begin{minipage}{\columnwidth}
\includegraphics[width=\textwidth]{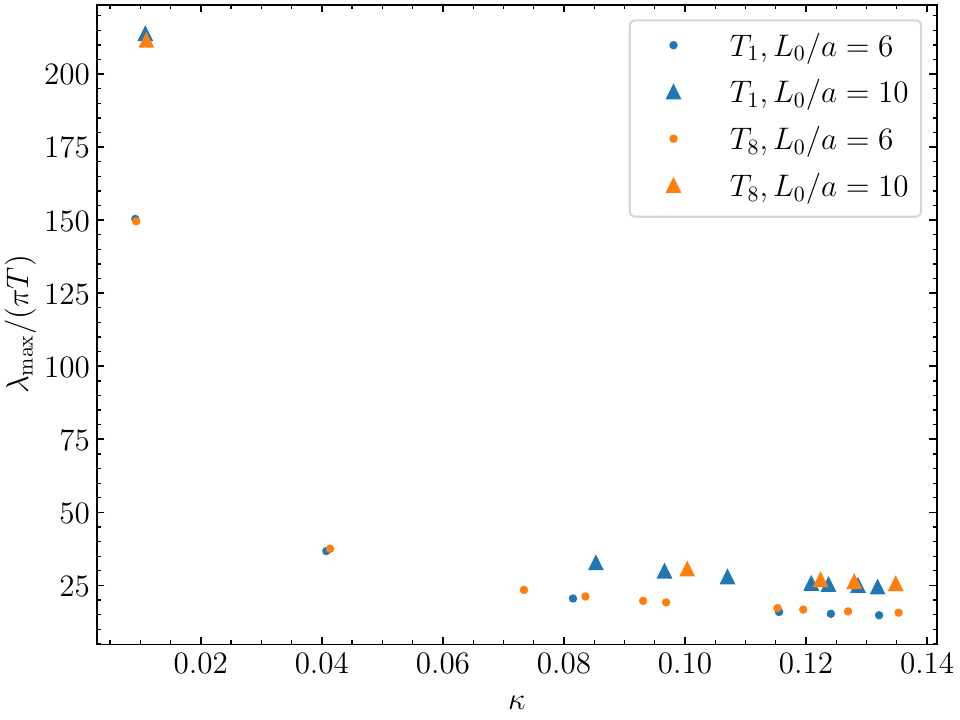}
\end{minipage}
\caption{Smallest (left plot) and largest (right plot) measured eigenvalues of the operator $\sqrt{D^\dagger D}$ as 
         functions of the hopping parameter, at some representative values of temperature and lattice spacing.}
\label{fig:Dirac_spectrum}
\end{center}
\end{figure*}

\subsection{Tuning of the HMC}
\label{ssec:Tuning of the HMC}
We performed the tuning of the algorithm parameters at the different values of the hopping parameter $\kappa$, required by the integration in the quark mass, 
on $6\times48^3$ lattices at one representative value of bare coupling and shift $L_0\vxi = (8a,0,0)$.
We started from the ensemble at the lowest value of $\kappa$ and we progressively increased it towards the 
critical value, exploring several algorithmic setups at each value of the hopping parameter.
We defined a performance estimator given by the ratio of the computational time over acceptance rate: the best algorithm 
at given $\kappa$ is the one that minimizes this estimator (with a lower bound of about $90\%$ on the acceptance).
We also monitored the spectrum of the Dirac operator so as to optimally tune the twisted-masses for the Hasenbusch 
preconditioning and the RHMC parameters.

At the smallest values of hopping parameter we used a 2-level multistep algorithm where the gauge force is at the finest level 
and the fermionic forces are at the coarsest one~\cite{Sexton:1992nu}. 
The latter include the contributions from the up-down doublet and from the RHMC for the strange quark.
Both levels have been integrated with a 4th order Omelyan-Mryglod-Folk (OMF) scheme~\cite{Omelyan:2002qkh},
the finest with 1 step and the coarsest with a tunable number of steps ranging between 7 and 9 for increasing hopping
parameter and decreasing temperature. 
By looking at the performance estimator introduced above and at the acceptance, we observed that the Hasenbusch 
preconditioning of the up-down doublet determinant is not required up to $\kappa\sim0.10$.
In the interval $0.10\lesssim\kappa\lesssim0.12$ this simple algorithm is refined with one Hasenbusch twisted-mass 
equal to $\sqrt{\lambda_{\rm min}\lambda_{\rm max}}$~\cite{Hasenbusch:2002ai}, where $\lambda_{\rm min}$ and $\lambda_{\rm max}$
are respectively the smallest and largest eigenvalues of the operator $\sqrt{D^\dagger D}$, whose determination is described later.
In this interval the number of steps for the molecular dynamics at the coarsest level ranges between 9 and 15.
For $\kappa\gtrsim0.12$, we opted for a frequency splitting of the RHMC in two terms, and we added a third integration 
level to profit from the hierarchy in magnitude of these two contributions.
The two finest levels have been integrated with 1 step of 4th order OMF scheme, and the coarsest level with 
a 2nd order OMF scheme with a tunable number of steps from 12 to 17.
More elaborated algorithms were not competitive up to $\kappa\gtrsim0.126$, where we split the up-down doublet determinant 
with one further Hasenbusch twisted-mass.
We tried many setups, including a generalization of the rule proposed in Ref.~\cite{Hasenbusch:2002ai} to the two twisted-mass case.
At the end we found that the two values $\lambda_{\rm min}+\frac{1}{3}(\lambda_{\rm max}-\lambda_{\rm min})$ 
and $\lambda_{\rm max}-\frac{1}{3}(\lambda_{\rm max}-\lambda_{\rm min})$ are a good choice in terms of performance.
Here the number of steps for the molecular dynamics at the coarsest level ranges between 13 and 20.

At a given $L_0/a$ and $g_0^2$ we have measured $\lambda_{\rm min}$ and $\lambda_{\rm max}$ on the thermalization ensembles 
at several values of the hopping parameter and shift $\vxi=(1+2a/L_0,0,0)$.
Figure~\ref{fig:Dirac_spectrum} shows the lowest eigenvalue (left plot) and the largest one (right plot) 
as functions of $\kappa$ for the temporal sizes $L_0/a=6,10$ and at the bare parameters of the temperatures $T_1$, $T_8$.

We have used these results also for the tuning of the rational approximation interval $[r_a,r_b]$ in the RHMC.
At given bare parameters we have chosen the conservative values $r_a=0.7\lambda_{\rm min}$ and $r_b = 1.2\lambda_{\rm max}$
to ensure that the spectrum of the Dirac operator is safely contained in the approximation interval.
The number of poles ranges between 3 and 10 as $\lambda_{\rm min}$ decreases, and has been chosen so that the systematic error 
of the rational approximation is completely negligible within the final accuracy of the numerical results of the simulations~\cite{openQCD,Luscher:2012av}.

\subsection{Generation of ensembles}
\label{ssec:Genration of ensembles}
At each set of bare parameters $\{L_0/a$, $g_0^2$, $\kappa$, $\vxi\}$ the gauge configurations were first generated on lattices with spatial 
sizes $L/a=48$, with a statistics of $500$-$1000$ trajectories of $2$ MDUs each.
We have proceeded from high to low temperatures, using the last configuration of each ensemble as 
starting point for the ensemble at the same shift and closest hopping parameter at the next lower temperature.
These small volume ensembles were used to test the performance of the algorithms, to measure the spectrum of the Dirac 
operator and to provide thermalization for the $L/a=144$ ensembles.
We have always monitored the components of the energy-momentum tensor and the $\mathbb{Z}_3$ phase of 
the generated configurations through the Polyakov loop. 
We have also monitored the Monte Carlo history of the average plaquette and of the topological charge at Wilson flow 
time $t_{\rm wf}$ fixed by the condition $T\sqrt{8t_{\rm wf}}=1/\sqrt{10}\approx0.3$~\cite{Fritzsch:2013yxa,DallaBrida:2016kgh}.

The lattices with target spatial size $L/a=144$ were generated by tripling the $L/a=48$ ones in the spatial directions, 
after verifying that the initial smaller-volume configuration was in the zero topological sector and had trivial $\mathbb{Z}_3$ phase.
At the temperature $T_8$, resolution $L_0/a=6$ and $\kappa\lesssim0.10$, we further refined the thermalization procedure 
considering two steps: from $L/a=24$ lattices to $L/a=72$ ones and finally to the target volume by anti-periodically 
extending the $L/a=72$ lattices to the $L/a=144$ ones so as to start with exactly zero topology at the target volume.
The Monte Carlo history of the average plaquette suggested to skip $50$-$100$ trajectories as 
thermalization, after which we finally measured our observables.
The same checks on energy-momentum tensor, topology and $\mathbb{Z}_3$ phase have been performed 
on the large volume ensembles as well.

\section{Scalar density variance reduction}
\label{app:Scalar density variance reduction}
The scalar density is the primary observable for the computation of $\dynamic$.
To optimize our numerical approach it is important to understand its behaviour, and the one
of its variance, in the limit of large values of the bare quark mass, i.e. their hopping parameter expansion.

\subsection{Random sources}
On a given gauge configuration, we estimate stochastically the trace of the quark propagator
using U$(1)$ random sources $\etabar,\eta$ ~\cite{Bitar:1988bb,Dong:1993pk,Michael:1998sg}, whose 2-point function is 
\begin{equation}
    \corr{\etabar_\alpha^a(x)\eta_\beta^b(y)} = \delta_{xy}\delta^{ab}\delta_{\albe}\,,
    \label{eq:eta_2pf}
\end{equation}
being $x,y$ lattice sites, $a,b$ colour indices and $\alpha,\beta$ spin indices.
Using this property it is immediate to show that
\begin{equation}
    \cO[U,\eta] = \frac{1}{N_s}\sum_{i=1}^{N_s} \frac{a^4}{V} \sum_x\etabar_i(x)\left\{S[U]\eta_i\right\}(x)
    \label{eq:estimator_O}
\end{equation}
is an estimator for the trace of the (single flavour) quark propagator $S=(D+m_0)^{-1}$.
In this equation $V=L_0L^3$ is the volume of the lattice and $\etabar_i,\eta_i$ are $N_s$ independent random 
sources (spin and colour indices have been omitted).
Thus the expectation value of the scalar density can be written as follows,
\begin{equation}
    a^3\corr{\psibar\psi} = -\frac{a^3}{V}\Nf\corr{\Tr{S}} = -a^{-1}\Nf\corr{\cO}\,,
    \label{eq:Sd_TrS}
\end{equation}
where the trace $\Tr{\cdot}$ is over spacetime, spin and colour indices and $N_f=3$ in our case.
The variance of $\cO$ reads
\begin{multline}
    \frac{V^2}{a^8}\left(\corr{\cO^2} - \corr{\cO}^2\right) = \Big[ \corr{{\rm Tr}^2 \{S\}} - \corr{\Tr{S}}^2 \Big] \\
    + \frac{1}{N_s} \Big[ \corr{\Tr{S^2}} - \langle\sum_{x,c,\sigma} \left[S_{\sigma\sigma}^{cc}(x,x)\right]^2\rangle \Big]\,,
    \label{eq:varO}
\end{multline}
where the term in the first square brackets is the gauge noise, while the term in the second square brackets 
is the additional noise due to the use of random sources.
The computation of Eq.~\eqref{eq:varO} involves a 4-point function of the random sources, 
whose expression for U$(1)$ fields is
\begin{multline}
    \corr{\etabar_\alpha^a(x)\eta_\beta^b(y)\etabar_\gamma^c(z)\eta_\delta^d(w)} =
    \delta_{xy}\,\delta_{zw}\,\delta^{ab}\,\delta^{cd}\,\delta_{\alpha\beta}\,\delta_{\gamma\delta} \\
    + \delta_{xw}\,\delta_{yz}\,\delta^{ad}\,\delta^{bc}\,\delta_{\alpha\delta}\,\delta_{\beta\gamma}
    - \delta_{xyzw}\,\delta^{abcd}\,\delta_{\alpha\beta\gamma\delta}\,.
\end{multline}

\subsection{Hopping parameter expansion}
To study the large mass behaviour it is convenient to rewrite the operator $D+m_0$ as follows, 
\begin{equation}
    D + m_0 = H + \frac{1}{2a\kappa}\,, \quad H \equiv D - \frac{4}{a}\,,
\end{equation}
where $H$ includes the hopping terms of the Wilson-Dirac operator $D_{\rm w}$ and the improvement operator $aD_{\rm sw}$, and 
$\kappa = 1/(8+2am_0)$ is the hopping parameter.
In the $\kappa\to0$ expansion, the quark propagator reads
\begin{equation}
    a^{-1}S = \frac{2\kappa}{1+2\kappa aH} = 2\kappa\sum_{n=0}^\infty\left(-2\kappa aH\right)^n\,.
    \label{eq:S_hpe}
\end{equation}
We are interested in the computation of the shift derivative of  the expectation value of the scalar density, see Eq.~\eqref{eq:Df_quark}.
Its hopping expansion at leading order is
\begin{equation}
    a^3\frac{\Delta\corr{\psibar\psi}_\vxi}{\Delta\xi_k} = 
    -8\kappa^3\frac{a^4}{V}\Nf\frac{\Delta\corr{T_2}_{\vxi}}{\Delta\xi_k} + O(\kappa^4)\,,
    \label{eq:DSdDxi_hpe}
\end{equation}
where we introduced the convenient notation
\begin{equation}
    T_n\equiv \Tr{(aH)^n} = a^n\sum_{x,c,\sigma} (H^n)_{\sigma\sigma}^{cc}(x,x)\,.
\end{equation}
On the other hand, the hopping expansion of the variance of the estimator $\cO$ is
\begin{multline}
   \frac{V^2}{a^{10}}\left(\corr{\cO^2} - \corr{\cO}^2\right) = (2\kappa)^6\Big[\corr{T_2^2}-\corr{T_2}^2 + O(\kappa)\Big] \\
   + \frac{1}{N_s} (2\kappa)^4\Big[\corr{T_2} - \langle\sum_{x,c,\sigma} \left[aH_{\sigma\sigma}^{cc}(x,x)\right]^2\rangle + O(\kappa)\Big]\,,
   \label{eq:varO_hpe}
\end{multline}
where the $O(\kappa^6)$ contribution comes from the fluctuations of the gauge configurations only, 
while the $O(\kappa^4)$ contribution originates from the presence of random sources.
Therefore, the leading scaling of the overall variance in the hopping expansion is of $O(\kappa^4)$.
As a consequence, the signal-to-noise ratio of the quantity in Eq.~\eqref{eq:DSdDxi_hpe} is of $O(\kappa)$ and decreases 
for increasing quark mass values.
The slower decreasing of the $O(\kappa^4)$ contribution in Eq.~\eqref{eq:varO_hpe} may be compensated by increasing 
the number of sources as $N_s\propto\kappa^{-2}$, although this would make the computation more and more demanding as $\kappa\to0$.

\subsection{Hopping subtraction}
\begin{figure}[ht]
    \centering
    \includegraphics[width=\columnwidth]{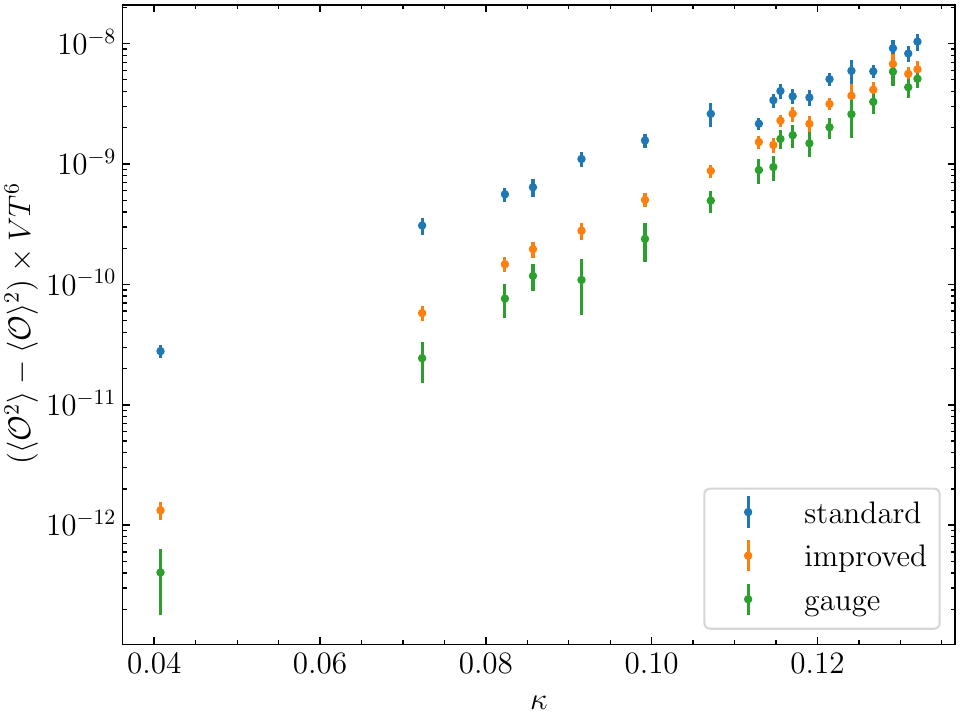}
    \caption{Variance computed with the estimators $\cO$ (blue points) and $\wt\cO$ (orange points), 
    respectively defined in Eqs.~\eqref{eq:estimator_O} and~\eqref{eq:estimator_O_impr}.
    For comparison, the gauge contribution is also shown (green points).
    These data are a representative case obtained from $6\times144^3$ lattices at inverse bare coupling $6/g_0^2=8.5403$.}
    \label{fig:variance_vs_kappa}
\end{figure}
Following Ref.~\cite{Giusti:2019kff} we have introduced an improved version of the estimator, called $\wt\cO$, obtained by replacing
\begin{equation}
    S \to S + 4\kappa^2a^2H
    \label{eq:estimator_O_impr}
\end{equation}
in the definition Eq.~\eqref{eq:estimator_O}, i.e. the quark propagator is subtracted by its leading non-trivial order 
in the hopping parameter expansion.
Since $H$ is traceless, the expectation values of the original estimator and of the improved one coincide: $\corr{\cO} = \corr{\wt\cO}$.
The subtraction removes the leading $O(\kappa^4)$ terms in the hopping expansion of the variance 
so that the total variance is now of $O(\kappa^6)$,
\begin{multline}
   \frac{V^2}{a^{10}}\left(\corr{\wt\cO^2} - \corr{\wt\cO}^2\right) = (2\kappa)^6\Big[\corr{T_2^2}-\corr{T_2}^2 \\
   + \frac{1}{N_s} \Big(\corr{T_4} - \langle\sum_{x,c,\sigma}[(a^2H^2)^{cc}_{\sigma\sigma}(x,x)]^2\rangle\Big) \Big] \\
   + O(\kappa^7)\,,
   \label{varO_impr_hpe}
\end{multline}
and the signal-to-noise ratio of the derivative in Eq.~\eqref{eq:DSdDxi_hpe} is consequently of $O(1)$ in the hopping parameter.
The effect of the improvement can be seen in Figure~\ref{fig:variance_vs_kappa}, where the comparison of the variance computed 
with the standard estimator in Eq.~\eqref{eq:estimator_O} and with the improved one through Eq.~\eqref{eq:estimator_O_impr} is shown, 
at several values of the hopping parameter.
The variance of the improved estimator is smaller than the one of the standard definition at all values of $\kappa$ considered,
and for $\kappa\to0$ the former decreases faster than the latter with a trend very similar to the one of the gauge noise component,
which is shown in the plot as well.

This simple subtraction leads to a factor up to 2.5 gain in the statistical error of $\dynamic$ 
with negligible additional computational cost, as the effort of evaluating numerically the estimators $\wt\cO$
and $\cO$ on a given gauge configuration is by far dominated by the computation of the quark propagator.

\section{Optimization of the statistics}
\label{sec:Optimization_of_the_statistics}

\begin{table}
\centering
\begin{tabular}{|c|c|c|c|c|}
    \hline
     $k$  & $L_0/a=4$ & $L_0/a=6$ & $L_0/a=8$ & $L_0/a=10$ \\
    \hline
     1  &   50      &   100     &   50      &    50        \\
     2  &   50      &   100     &  150      &   100        \\
     3  &   50      &   100     &  150      &   100         \\
     4  &   50      &   100     &  200      &   200         \\
     5  &   50      &   100     &  150      &   250         \\
     6  &   50      &   100     &  150      &   250         \\
     7  &   50      &   100     &  150      &   200         \\
     8  &   50      &   100     &  150      &   150         \\
     9  &   50      &   100     &  100      &   100         \\
    10  &   50      &   100     &   50      &    50        \\
    \hline
    11  &   50      &   100     &  250      &    300        \\
    12  &   50      &   100     &  400      &    600        \\
    13  &   50      &   100     &  450      &    700        \\
    14  &   50      &   100     &  400      &    700        \\
    15  &   50      &   100     &  300      &    500        \\
    16  &   50      &   100     &  150      &    350        \\
    17  &   50      &   100     &  150      &    150        \\
    \hline
    18  &   50      &   100     &  1400     &   2400         \\
    19  &   50      &   100     &  1200     &   2100         \\
    20  &   50      &   100     &   150     &    300        \\
    \hline
\end{tabular}
\caption{Number of measurements for each $L_0/a$ optimized to obtain 
$\frac{\Delta(f_\vxi-f_\vxi^\infty)}{\Delta\xi_k}$ with a $0.5\%$ relative error
for $L_0/a=4,6,8$ and a $1.0\%$ relative error for $L_0/a=10$, at the temperature $T_1$ and $L/a=144$. 
Each row corresponds to one of the Gauss quadrature points for the numerical determination
of the integral in the bare quark mass, and they are labelled by the index $k$ as in Eq.~\eqref{eq:totvar_dynamic_contr}.
The bare subtracted quark mass increases from top to bottom.
Horizontal lines separate the three integration domains as described in Eq.~\eqref{eq:integral_mq_split}.}
\label{tab:ntr_opt}
\end{table}
We discuss here the strategy that we have implemented for the optimization of the statistics needed for the 
numerical determination of $\dynamic$, computed as described in Section~\ref{sec:Numerical computation}.
At a given set of bare parameters $L_0/a$ and $g_0^2$ the error squared of this contribution is propagated from the 
fluctuations of the scalar density as follows,
\begin{equation}
    \sigma^2\left[\frac{1}{T^4}\frac{\Delta(f_\vxi-f_\vxi^\infty)}{\Delta\xi_k}\right]
    = \sum_{k=1}^{n_G}\omega_k^2\,\sigma^2\left[\frac{1}{T^3}\frac{\Delta\corr{\psibar\psi}_\vxi^{m_{q,k}}}
    {\Delta\xi_k}\right]\,,
    \label{eq:totvar_dynamic_contr}
\end{equation}
where the index $k$ runs over the $n_G=20$ Gauss points prescribed by the integration scheme of Eq.~\eqref{eq:integral_mq_split}, 
$\omega_k$ are the related Gauss weights and on the right it appears the error squared of the discrete derivative of 
$\corr{\psibar\psi}_\vxi^{m_q}$ with respect to the shift. 
The latter is obtained as the sum in quadrature of the errors at the same bare quark mass and two shifts.
The contribution to the final error of the different Gauss points is thus modulated by the square of the Gauss weights $\omega_k^2$, 
whose magnitude is maximum in the middle of the integration intervals and decreases symmetrically towards the boundaries.
This suggests that the number of measurements of the scalar density at different bare quark masses can be tuned so as to 
minimize the computational cost for attaining a given target accuracy.
In this optimization we also considered that the computational cost of simulations and the variance of the observable depend
on the bare quark mass.

Therefore, at fixed $L_0/a$ and $g_0^2$ the quantity to be minimized is the total computational cost, under the constraint that the
final relative error on $\dynamic$ is equal to a target value. 
This problem can be solved through the method of Lagrange multipliers, the variables to be optimized being the number of measurements
for the determination of $\corr{\psibar\psi}_\vxi^{m_q}$ at the different Gauss points.
We have chosen a target relative accuracy of $0.5\%$ at $L_0/a=4,6,8$ and $1.0\%$ at $L_0/a=10$.
As an example, the optimized results for the temperature $T_1$ are reported in Table~\ref{tab:ntr_opt}.
At the other temperatures, we increased these optimized numbers by a monotonically growing factor up to 2 at the lowest temperature.
The quoted numbers of measurements refer to both shifts.
For the coarsest resolutions $L_0/a=4,6$ there was little gain from this 
procedure, thus we have chosen a fixed number of measurements for all the Gauss quadrature points. 
We have used the optimization for the finest lattices $L_0/a=8,10$ which are the most expensive to be generated. 
The gain in computational cost is up to a factor $2$ in comparison with the case where the same number
of measurements is chosen for all the Gauss quadrature points, at the same fixed target accuracy.
We have also generalized this procedure so as to optimize the number of sampled noise sources 
(see Appendix~\ref{app:Scalar density variance reduction}) as well,
but the latter turned out to be mostly constant and for convenience we have fixed it to $100$ for all the simulations.

\section{Further numerical results}
\label{app:Further numerical results}
In this Appendix we collect some auxiliary numerical results which complement the ones given in the main text.

\subsection{Results for \mtht{$\frac{\Delta\corr{\overline{S_G}}^\infty_\vxi}{\Delta\xi_k}$}}
As a representative case, we give in Table~\ref{tab:DSGDxi} the numerical results for the derivative in the shift of the expectation 
value of the density of the Wilson plaquette action, defined in Eq.~\eqref{eq:SG}, computed in pure gauge simulations at the lattice 
resolution $L_0/a=6$ and at the values of bare coupling prescribed by the integration scheme presented in Section~\ref{sec:Numerical computation}.

\begin{table}[t]
    \centering
    \begin{tabular}{|c|c||c|c|}
    \hline
    & & & \\[-0.25cm]
    $6/g_0^2$ & $\frac{\Delta\corr{\overline{S_G}}^\infty_\vxi}{\Delta\xi_k} \times 10^{4}$ &
    $6/g_0^2$ & $\frac{\Delta\corr{\overline{S_G}}^\infty_\vxi}{\Delta\xi_k} \times 10^{4}$ \\
  	& & & \\[-0.25cm]
    \hline
    15.0000 &    0.22(3) &   7.5523 &    2.71(9) \\
    \cline{1-2}
    13.9517 &    0.21(5) &   7.4227 &    3.15(7) \\
    11.2500 &    0.32(5) &   7.2975 &    3.76(9) \\ 
    \cline{3-4}
     9.4249 &    0.56(7) &   7.2245 &    4.13(8) \\ 
     \cline{1-2}
     8.9975 &    0.83(6) &   7.0990 &    4.95(6) \\ 
     \cline{1-2}
     8.9457 &    0.84(6) &   6.9778 &    5.80(9) \\ 
     \cline{3-4}
     8.7641 &    0.97(5) &   6.9267 &    6.41(9) \\
     8.5897 &    1.12(5) &   6.8622 &    7.16(6) \\ 
     \cline{1-2}
     8.5026 &    1.12(4) &   6.7699 &    8.46(8) \\
     8.3755 &    1.24(4) &   6.6801 &  10.24(17) \\
     8.2522 &    1.44(6) &   6.6201 &  11.61(14) \\ 
     \cline{1-4}
     8.1811 &    1.49(5) &   6.5887 &  12.49(25) \\
     8.0601 &    1.69(5) &   6.5254 &  14.55(29) \\
     7.9426 &    1.78(5) &   6.4350 &    18.9(4) \\ 
     \cline{1-2}
     7.8719 &    1.96(7) &   6.3470 &    25.9(5) \\
     7.7467 &    2.23(6) &   6.2883 &    34.5(7) \\
     7.6255 &    2.56(9) &          &            \\
    \hline
    \end{tabular}
    \caption{Numerical results for $\frac{\Delta\corr{\overline{S_G}}^\infty_\vxi}{\Delta\xi_k}$ computed on $6\times144^3$ lattices
    at the given values of inverse bare coupling $6/g_0^2$.
    Horizontal lines indicate the grouping of bare couplings according to the
    integration scheme in Table~\ref{tab:tab_g02_integral}.} 
    \label{tab:DSGDxi}
\end{table}

\subsection{Results for \mtht{$\frac{\Delta\corr{\psibar\psi}_\vxi^{m_q}}{\Delta\xi_k}$}}
As a representative case we collect in Table~\ref{tab:DSdDxi} the results for the derivative in the shift of the expectation 
value of the scalar density defined in Eq.~\eqref{eq:Sd_TrS} computed in lattice QCD simulations with $L_0/a=6$ and $6/g_0^2=8.5403$ 
(corresponding to the temperature $T_1$), at the hopping parameter values prescribed by the integration scheme 
presented in Section~\ref{sec:Numerical computation}.

\begin{table}[t]
    \centering
    \begin{tabular}{|c|c||c|c|}
    \hline
    & & & \\[-0.25cm]
    $\kappa$ & $\frac{\Delta\corr{\psibar\psi}_\vxi}{\Delta\xi_k} \times 10^3$ &
    $\kappa$ & $\frac{\Delta\corr{\psibar\psi}_\vxi}{\Delta\xi_k} \times 10^3$ \\
    & & & \\[-0.25cm]
    \hline
      $0.132067$ & $-0.491(14)$ & $0.112938$ & $-0.770(7)$ \\
      $0.130958$ & $-1.457(13)$ & $0.107136$ & $-0.3191(29)$ \\
      $0.129108$ & $-2.608(13)$ & $0.099197$ & $-0.0962(28)$ \\
      $0.126736$ & $-3.132(10)$ & $0.091536$ & $-0.0279(21)$ \\
      $0.124100$ & $-2.888(8)$ & $0.085678$ & $-0.0133(15)$ \\
      $0.121455$ & $-2.299(8)$ & $0.082296$ & $-0.0073(16)$ \\
      \cline{3-4}
      $0.119031$ & $-1.757(8)$ & $0.072311$ & $-0.0006(8)$ \\
      $0.117012$ & $-1.346(6)$ & $0.040748$ & $-0.00015(11)$ \\
      $0.115533$ & $-1.115(6)$ & $0.009185$ & $-0.0000001(14)$ \\
      $0.114683$ & $-0.990(5)$ &            &                 \\
    \hline
    \end{tabular}
    \caption{Numerical results for $\frac{\Delta\corr{\psibar\psi}_\vxi}{\Delta\xi_k}$ computed on $6\times144^3$ lattices
    at inverse bare coupling $6/g_0^2=8.5403$ and at the given values of hopping parameter $\kappa$.
    Horizontal lines indicate the three sets of 10, 6 and 3 Gauss points according to the
    integration scheme described in Subsection~\ref{ssec:Dynamic contribution}.}
    \label{tab:DSdDxi}
\end{table}

\subsection{Covariance of continuum limit results}
We report in Table~\ref{tab:cov_cicj} the entries of the covariance matrix of the continuum limits $c_i$, $i=0,...,8$ of the best fit \texttt{id3} 
in Table~\ref{tab:fits_comparison}, each normalized to the product of the errors. 
We also give the normalized covariance values of the continuum limits $c_i$ with the coefficient $d_{23}$
which parameterizes the discretization effects:
\begin{align}
    \frac{{\rm cov}(c_i,d_{23})}{\sigma(c_i)\sigma(d_{23})} =
     (& -0.613, -0.683, -0.701, \nonumber \\ 
          & -0.702, -0.716, -0.721, \nonumber \\ 
          & -0.754, -0.731, -0.737)\,.
    \label{eq:cov_cid23}
\end{align}

\begin{table}[t]
   \centering
   \begin{tabular}{|cc|cc|cc|}
   \hline
   & & & & & \\[-0.25cm]
   $(i,j)$ & $\frac{{\rm cov}(c_i,c_j)}{\sigma(c_i)\sigma(c_j)}$ &
   $(i,j)$ & $\frac{{\rm cov}(c_i,c_j)}{\sigma(c_i)\sigma(c_j)}$ &
   $(i,j)$ & $\frac{{\rm cov}(c_i,c_j)}{\sigma(c_i)\sigma(c_j)}$ \\
   & & & & & \\[-0.25cm]
   \hline
   $(0,1)$ & $0.687$ & $(1,6)$ & $0.775$ & $(3,7)$ & $0.765$ \\
   $(0,2)$ & $0.690$ & $(1,7)$ & $0.755$ & $(3,8)$ & $0.763$ \\
   $(0,3)$ & $0.688$ & $(1,8)$ & $0.753$ & $(4,5)$ & $0.774$ \\
   $(0,4)$ & $0.696$ & $(2,3)$ & $0.757$ & $(4,6)$ & $0.796$ \\
   $(0,5)$ & $0.696$ & $(2,4)$ & $0.766$ & $(4,7)$ & $0.775$ \\
   $(0,6)$ & $0.707$ & $(2,5)$ & $0.766$ & $(4,8)$ & $0.774$ \\
   $(0,7)$ & $0.700$ & $(2,6)$ & $0.787$ & $(5,6)$ & $0.797$ \\
   $(0,8)$ & $0.696$ & $(2,7)$ & $0.766$ & $(5,7)$ & $0.776$ \\
   $(1,2)$ & $0.750$ & $(2,8)$ & $0.763$ & $(5,8)$ & $0.775$ \\
   $(1,3)$ & $0.747$ & $(3,4)$ & $0.764$ & $(6,7)$ & $0.799$ \\
   $(1,4)$ & $0.756$ & $(3,5)$ & $0.764$ & $(6,8)$ & $0.798$ \\
   $(1,5)$ & $0.755$ & $(3,6)$ & $0.785$ & $(7,8)$ & $0.781$ \\
   \hline
   \end{tabular}
   \caption{Entries of the covariance matrix cov$(c_i,c_j)$ of the continuum results of the best fit \texttt{id3} 
            in Table~\ref{tab:fits_comparison}, normalized to the product of the errors $\sigma(c_i)$ and $\sigma(c_j)$.}
   \label{tab:cov_cicj}
\end{table}

\begin{figure*}[t]
   \begin{center}
   \begin{minipage}{\columnwidth}
   \includegraphics[width=\textwidth]{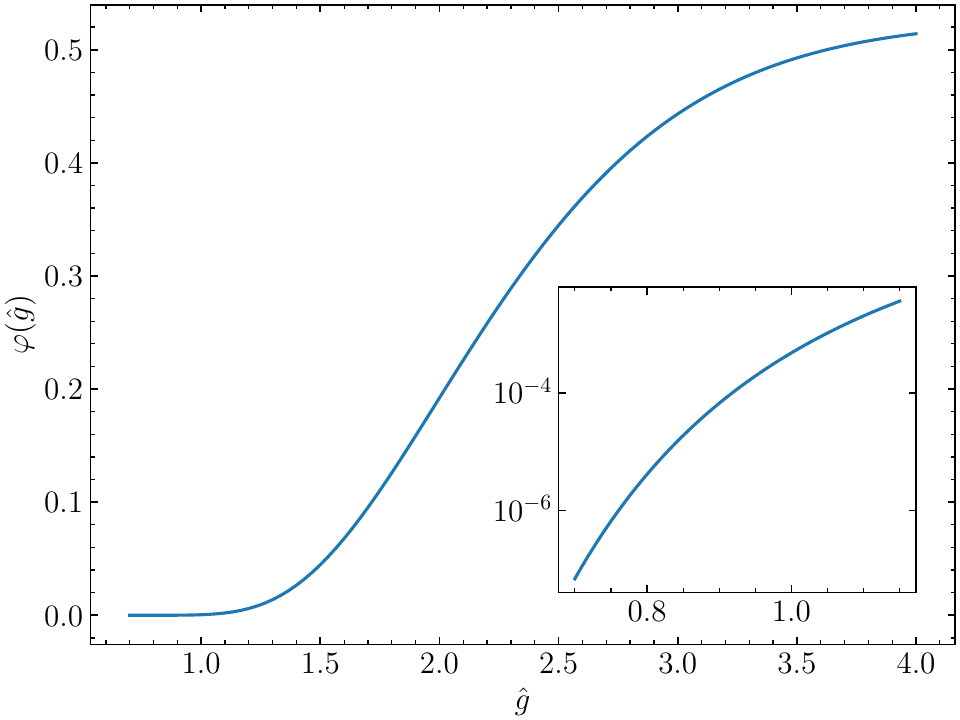}
   \end{minipage}%
   \begin{minipage}{\columnwidth}
   \includegraphics[width=\textwidth]{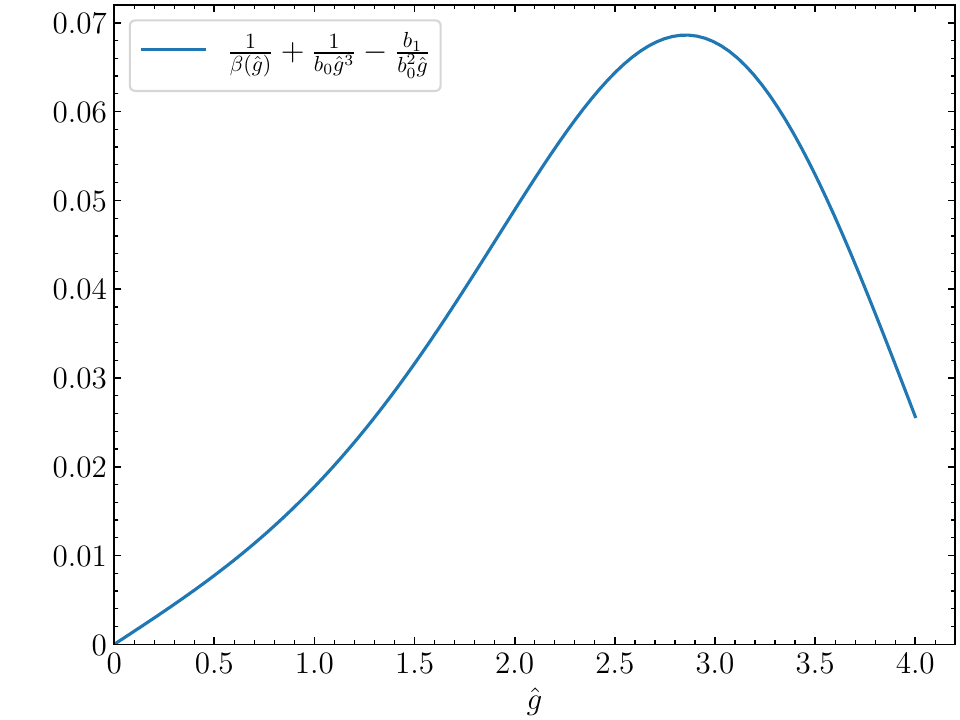}
   \end{minipage}
   \end{center}
   \caption{
   Left: plot of the function $\varphi$ in Eq.~\eqref{eq:phi}. The smaller plot is a detail for the interval 
   $0.8\lesssim\hat g\lesssim 1.1$, with log scale on the $y$-axis.
   Right: plot of the integrand function of ${\cal I}_\beta(\hat g)$ in Eq.~\eqref{eq:integral}.
   }
   \label{fig:integrand_phi}
\end{figure*}

\begin{table}[th]
	\centering
	\begin{tabular}{|c|c|}
		\hline
		$\mu/\Lambda_\MSbar$ & $\hat g(\mu/\Lambda_\MSbar)$ \\
		\hline
                $2$ & $3.62600$ \\
		$10^1$ & $1.71582$ \\
		$10^2$ & $1.25818$ \\
		$10^3$ & $1.04760$ \\
		$10^4$ & $0.91833$ \\
		$10^5$ & $0.82827$ \\
		\hline
	\end{tabular}
	\caption{Results for the running coupling $\hat g$ at some values of the renormalization scale.}
	\label{tab:results}
\end{table}

\section{Evaluation of the $\MSbar$ strong coupling at five-loop order}
\label{app:Strong coupling at five-loop in MSbar scheme}
In this Appendix we collect the details for the computation of the renormalized coupling $\hat g$, appearing in 
Sections~\ref{sec:Entropy density} and~\ref{sec:Equation of State}.
It is defined as the five-loop $\MSbar$ coupling for QCD with $N_f=3$ flavours,
evaluated at the renormalization scale $\mu=2\pi T$.
The $\Lambda$-parameter in this scheme is defined by~\cite{DallaBrida:2018rfy}
\begin{equation}
	\Lambda_\MSbar = \mu \varphi(\hat g(\mu/\Lambda_\MSbar))\,,
	\label{eq:Lambda}
\end{equation}
where the function $\varphi(\hat g)$ is
\begin{equation}
	\varphi(\hat g) = \left(b_0 \hat g^2\right)^{-b_1/(2b_0^2)} e^{-1/(2b_0\hat g^2)} e^{-{\cal I}_\beta(\hat g)}\,
	\label{eq:phi}
\end{equation}
with
\begin{equation}
	{\cal I}_\beta(\hat g) = \int_0^{\hat g}dg\left[\frac{1}{\beta(g)} + \frac{1}{b_0g^3} - \frac{b_1}{b_0^2 g}\right]\,,
	\label{eq:integral}
\end{equation}
and the $\beta$-function is expanded in perturbation theory as follows:
\begin{equation}
	\beta(\hat g) = \mu \frac{\partial \hat g}{\partial \mu} \xrightarrow{\hat g\to 0} -\hat g^3 \sum_{k\geq0} b_k \hat{g}^{2k}\,.
	\label{eq:beta_function}
\end{equation}
The perturbative coefficients $b_k$ are related to the coefficients $\beta_k$, $k=0,1,2,3,4$, determined in Ref.~\cite{Baikov:2016tgj},
by the relation
\begin{equation}
	b_k = \frac{\beta_k}{(2\pi)^{2k+2}}\,.
	\label{eq:bk_to_betak}
\end{equation}
We report here the values for the two lowest orders, which do not depend on the renormalization scheme:
\begin{equation}
	\beta_0 = \frac{1}{4}\left(11 - \frac{2}{3}N_f\right)\,, \,\,
	\beta_1 = \frac{1}{16}\left(102-\frac{38}{3}N_f\right)\,.
\end{equation}
At a given value of $\mu/\Lambda_{\MSbar}$, we have computed the corresponding coupling $\hat g(\mu/\Lambda_\MSbar)$
by solving numerically Eq.~\eqref{eq:Lambda}.
The evaluation of the function $\varphi(\hat g)$, represented in the left panel of Figure~\ref{fig:integrand_phi},
requires the computation of the integral ${\cal I}_\beta(\hat g)$.
To this purpose, we have used the optimized QUADPACK quadrature routines~\cite{piessens1983quadpack}
on the integrand function, plotted in the right panel of Figure~\ref{fig:integrand_phi}.
We have restricted the numerical approach to the conservative domain $2 \leq \mu/\Lambda_\MSbar \leq 10^5$, 
corresponding to $0.8 \lesssim \hat g \lesssim 3.6$.
This interval largely includes all the values of energy scale and renormalized coupling we are interested in.
Some representative results at several renormalization scales are reported in Table~\ref{tab:results}.

For $\mu/\Lambda_\MSbar > 10^5$, we have switched to the three-loop analytic formula obtained from Ref.~\cite{Deur:2016tte}.
The two couplings at the scale $\mu/\Lambda_\MSbar=10^5$ differ by less than $0.03\%$, and the discrepancy is expected to decrease 
as the renormalization scale increases.

\bibliography{main.bbl}

\end{document}